%% file: main.tex
\documentclass{lmcs}

\usepackage{lastpage}
\lmcsdoi{20}{4}{11}
\lmcsheading{}{\pageref{LastPage}}{}{}%
{Jun.~14,~2022}{Nov.~12,~2024}{}

\usepackage[utf8]{inputenc}

\makeatletter
\newcommand{\mathleft}{\@fleqntrue\@mathmargin0pt}
\newcommand{\mathcenter}{\@fleqnfalse}
\makeatother

\usepackage{amssymb,amsmath,amscd,amsbsy}
\usepackage{xspace}
\usepackage{graphicx,epic,eepic,gastex}
\usepackage{array}
\usepackage{hyperref}

\usepackage{booktabs}   
\usepackage{subcaption} 

\let\emptyset\varnothing
\let\epsilon\varepsilon

\newcommand\numberthis{\addtocounter{equation}{1}\tag{\theequation}}

\def\norm#1{\ensuremath{\lVert #1\rVert}}
\def\normInf#1{\ensuremath{\lVert #1\rVert_{\infty}}}
\def\abs#1{\ensuremath{\lvert #1\rvert}} 
\def\Abs#1{\ensuremath{\Big\lvert #1\Big\rvert}}

\newcommand{\nat}{\mathbb N} 
\newcommand{\rat}{{\mathbb Q}}

\newcommand{\real}{{\mathbb R}}

\newcommand{\tuple}[1]{\langle #1 \rangle}

\newcommand{\T}{\mathcal{T}}
\newcommand{\R}{\mathcal{R}}

\newcommand{\val}{\mathit{val}}

\newcommand{\transpose}{\intercal}

\newcommand{\Supp}{\mathsf{Supp}}
\renewcommand{\P}{{\mathbb P}}
\newcommand{\E}{{\mathbb E}}

\DeclareMathOperator{\lcm}{lcm}

\newcommand{\straa}{\sigma}

\newcommand{\MP}{\operatorname{\mathsf{MP}}}
\newcommand{\M}{\mathcal{M}}
\newcommand{\calE}{\mathcal{E}}
\newcommand{\reach}{\mathsf{reach}}
\newcommand{\reset}{\mathsf{reset}}
\newcommand{\dif}{\mathsf{dif}}
\newcommand{\ec}{\mathsf{ec}}
\newcommand{\trans}{\mathsf{trans}}

\begin{document}
\mathcenter

\title[Stochastic Processes with Expected Stopping Time]{Stochastic Processes with Expected Stopping Time}
\titlecomment{{\lsuper*}A preliminary version of this paper appeared 
in the \emph{Proceedings of the 36th Annual Symposium on Logic in Computer Science} (LICS), IEEE Computer Society Press, 2021~\cite{CD21}.}

\author[K. Chatterjee]{Krishnendu Chatterjee\lmcsorcid{0000-0002-4561-241X}}[a]
\author[L. Doyen]{Laurent Doyen\lmcsorcid{0000-0003-3714-6145}}[b]

\address{IST Austria}	
\email{krishnendu.chatterjee@ist.ac.at}  

\address{CNRS \& LMF, ENS Paris-Saclay, France}	
\email{doyen@lsv.fr}  

\begin{abstract}
Markov chains are the de facto finite-state model for stochastic dynamical systems,
and Markov decision processes (MDPs) extend Markov chains by incorporating non-deterministic
behaviors.
Given an MDP and rewards on states, a classical optimization criterion is 
the maximal expected total reward where the MDP stops after $T$ steps, which 
can be computed by a simple dynamic programming algorithm.
We consider a natural generalization of the problem where the stopping times 
can be chosen according to a probability distribution, such that the expected
stopping time is $T$, to optimize the expected total reward.
Quite surprisingly we establish inter-reducibility of the expected stopping-time 
problem for Markov chains with the Positivity problem (which is related to the well-known Skolem problem),
for which establishing either decidability or undecidability would be a major breakthrough.
Given the hardness of the exact problem, we consider the approximate version of the 
problem: we show that it can be solved in exponential time for Markov chains and in exponential space 
for MDPs.
\end{abstract}



\maketitle

\section{Introduction}\label{sec:introduction}

\noindent{\em Stochastic models and optimization.} 
The de facto model for stochastic dynamical systems is finite-state 
Markov chains~\cite{FV97,Gallager,Kemeny66}, 
with several application domains~\cite{BK08}. 
In modeling optimization problems, rewards are associated with states of the 
Markov chain, and the optimization criterion is formalized as the expected total reward 
provided that the Markov chain is stopped after $T$ steps~\cite{PT87,FV97}.
The extension of Markov chains to allow non-deterministic behavior gives rise to 
Markov decision processes (MDPs), and the optimization criterion is to maximize,
over all non-deterministic choices, the expected total reward for $T$ steps. 
This notion of optimization for fixed time is called {\em finite-horizon planning}, 
which has many applications in logic and verification~\cite{EMSS92,BCCSZ03} and 
control problems in artificial intelligence 
and robotics~\cite[Chapter~10, Chapter~25]{AIBook},~\cite[Chapter~6]{OR94}.

\smallskip\noindent{\em Optimization with expected stopping time.}
In the most basic case the stopping time for collecting rewards in the stochastic model
is a fixed constant~$T$. A natural generalization is to consider that the stochastic model 
can be stopped at a random time such that the expectation of the stopping time is $T$. 
We consider the problem of optimizing (maximizing/minimizing) the expected total 
reward, when the stopping-time probability distribution can be chosen arbitrarily 
such that the expected stopping time is $T$.
In other words, we consider stochastic models of Markov chains/MDPs 
with total reward, and instead of fixed stopping time $T$, 
we consider expected stopping time $T$.

\smallskip\noindent{\em Example and motivation.} 
Consider a classical example where a robot explores a region for natural resources 
(e.g., the well-studied RockSample problem in AI literature~\cite{SS04}), and the exploration of 
the robot is modeled as a Markov chain. 
The success of the exploration is characterized by the expected total reward,
and the stopping time $T$ denotes the expected duration of the exploration.  
The expected stopping-time problem asks to choose the probability distribution of 
the exploration duration to optimize the collected reward, satisfying the average 
exploration time.
A classical stopping-time distribution is the geometric distribution
where the stochastic model is stopped at every instant with probability~$\lambda$,
called \emph{discount factor}, which entails that the expected stopping time 
is $T = 1/\lambda$~\cite{FV97}.
The discount-factor model makes an assumption on the shape of the 
stopping-time distribution, whereas in realistic scenarios the distribution 
is not precisely known, or time-varying discount factors are considered~\cite{Dew12}. 
When the discount factors are not known, then robust solutions require the worst-case 
choice of the factors. Thus in many examples realistic modeling requires complex 
stopping-time distributions, and if the precise parameters are unknown, then 
a robust analysis requires to consider the worst-case stopping-time distribution.
Hence, when the stopping-time distribution is important yet unknown, 
a conservative estimate (i.e., lower bound) of the optimal value is obtained
using the worst-case choices. 
Thus we consider problems that represent robust extensions of the classical finite-horizon planning.

\smallskip\noindent{\em Previous and our results.}
For fixed stopping time $T$, the expected total reward for Markov chains and MDPs can be 
computed via a simple dynamic programming (or backward induction) 
approach~\cite[Chapter~4]{Puterman},~\cite{FV97, Howard60, BKNPS19}.
The optimization problem for Markov chains and MDPs with expected
stopping time has not been considered in the literature (to the best of 
our knowledge).
Our main results are as follows:
\begin{itemize}
\item In contrast to the simple algorithm for fixed stopping time $T$, we show that 
quite surprisingly the expected stopping-time problem is {\em Positivity}-hard.
The Positivity problem is known to be at least as hard as the well-known Skolem problem, 
whose decidability has been open for more than eight decades~\cite{OW14}.
Moreover, we establish inter-reducibility between the expected stopping-time problem 
and the Positivity problem, and thus show that for a simple variant 
(adding expectation to stopping time) of the classical Markov chain problem, 
establishing either decidability or undecidability would be a major breakthrough.

\item We then consider approximating the optimal expected total reward  
under the constraint that the expected stopping time is $T$, and show 
that for every additive absolute error $\epsilon>0$, the approximation can be achieved in time 
logarithmic in $1/\epsilon$ and exponential in the size of the Markov chain.

\item For MDPs we show that infinite-memory strategies are required. 
While the expected stopping-time problem is Positivity-hard for MDPs (since Markov chains are a 
special case), we show that the approximation problem can be solved in exponential space 
in the size of the MDP and logarithm of $1/\epsilon$.  

\end{itemize}

\smallskip\noindent{\em Comparison with related work.} 
The optimization problem with fixed expected stopping time has been considered 
for the simple model of graphs~\cite{CD19}, which is a model without stochastic aspects. 
The graph problem can be solved in polynomial time~\cite{CD19}, while 
in sharp contrast, we show that the problem is Positivity-hard for Markov chains. 

\begin{rem}\label{rmk:probabilistic-automata}
The expected stopping-time problem for Markov chains has a similar 
flavor as probabilistic automata (or blind MDPs)~\cite{RabinProb63}. 
In probabilistic automata a word (or letter sequence) must be provided
without the information about how the probabilistic automaton executes.
Similarly, for the expected stopping-time problem for Markov chains
the probability distribution for stopping times must be chosen
without knowing the execution of the Markov chain (in contrast to stopping criteria 
based on current state or accumulated reward, which rely on knowing the execution 
of the Markov chain).
For probabilistic automata, even for basic reachability, all problems
related to approximation are undecidable~\cite{MHC03}.
In contrast, we show that while the exact problem for expected stopping time
in Markov chains is Positivity-hard, the approximation problem can be 
solved in exponential time.
\end{rem}

\section{Preliminaries}\label{sec:prelim}

A \emph{stopping-time distribution} (or simply, a distribution) is a
function $\delta: \nat \to [0,1]$ such that $\sum_{t \in \nat} \delta(t) = 1$.
The support of $\delta$ is $\Supp(\delta) = \{t \in \nat \mid \delta(t) \neq 0\}$.
We denote by $\Delta$ the set of all stopping-time distributions, and by $\Delta^{\upuparrows}$
the set of all distributions $\delta$ with $\abs{\Supp(\delta)} \leq 2$,
called the \emph{bi-Dirac} distributions.

The \emph{expected utility} of a sequence $u = u_0, u_1, \dots$ of real numbers 
under a distribution $\delta$ is $\E_{\delta}(u) = \sum_{t \in \nat} u_t \cdot \delta(t)$.
In particular, the expected utility of the sequence $0,1,2,3,\dots$ of all natural numbers
is called the \emph{expected time} (of distribution $\delta$), denoted by~$\E_{\delta}$.


We recall the definition of the Positivity problem and of the related Skolem problem.
In the sequel, we denote by $M^t_{ij}$ the $(i,j)$ entry of the $t$-th power
of matrix $M$ (we should write it as $(M^t)_{i,j}$, but use this simpler
notation when no ambiguity can arise).

\medskip\noindent{\bf Positivity problem~\cite{OW14,AAOW15}.}
Given a square integer matrix $M$, 
decide whether there exists an integer $t \geq 1$ such that $M^t_{1,2} > 0$. 

\medskip\noindent{\bf Skolem problem~\cite{OW14,AAOW15}.}
Given a square integer matrix $M$, 
decide whether there exists an integer $t \geq 1$ such that $M^t_{1,2} = 0$. 
\medskip

The decidability of the Positivity and Skolem problems is a longstanding open question~\cite{OW14}, 
and there is a reduction from the Skolem problem to the Positivity problem
that increases the matrix dimension quadratically~\cite{HHH06,OW14}. 

\section{Markov Chains}
We present the basic definitions related to Markov chains
and the decision problems for the optimal total reward
with expected stopping time.

\subsection{Definitions}\label{sec:mc}
\noindent A \emph{Markov chain} is a tuple $\tuple{M,\mu,w}$ consisting of:
\begin{itemize}
\item an $n \times n$ stochastic matrix $M$
(in which all entries $M_{ij}$ are nonnegative rationals\footnote{For decidability and complexity results, we assume 
the numbers are rationals encoded as two binary numbers.}, 
and the sum $\sum_j M_{ij}$ of the elements in each row~$i$ is $1$),
\item an initial distribution $\mu \in ([0,1] \cap \rat)^n$ (viewed as
$1 \times n$ row vector, and such that $\sum_i \mu_i = 1$), and
\item a vector $w \in \rat^n$ of weights (or rewards). 
\end{itemize}

\noindent We also view $\mu$ and $w$ as functions $V \to \rat$ where $V = \{1,2,\dots,n\}$ 
is the set of vertices of the Markov chain. We often abbreviate Markov chains as $M$,
when $\mu$ and $w$ are clear from the context. We denote by $\norm{w} = \max_{v \in V} \abs{w(v)}$
the largest absolute value in $w$.

A Markov chain induces a probability measure on sequences of vertices of 
a fixed length, namely $\P(v_0 v_1 \dots v_k) = \mu(v_0) \cdot \prod_{i=0}^{k-1} M_{v_i,v_{i+1}}$.
Analogously, we denote by $\E(f)$ the expected value of   
the function $f: V^* \to \rat$ defined over finite sequences of vertices.

Given a stopping-time distribution $\delta: \nat \to [0,1]$, let $N_{\delta}$
be a random variable whose distribution is $\delta$. 
We are interested in computing the \emph{optimal} (worst-case) expected value (or simply the value)
of Markov chains with expected stopping time $T$, defined by:
\begin{align*}
\val(M,T) & = \inf_{\substack{\delta \in \Delta \\ \E_{\delta} = T}} \E\left[\sum_{i=0}^{N_{\delta}} w(v_i)\right] = \inf_{\substack{\delta \in \Delta \\ \E_{\delta} = T}} \E\left[\sum_{i=0}^{N_{\delta}} \mu \cdot M^i \cdot w^\transpose \right] \\
 & = \inf_{\substack{\delta \in \Delta \\ \E_{\delta} = T}} \sum_{t=0}^{\infty} \delta(t) \cdot u_t, \\
\end{align*}
where $w^\transpose$ is the transpose of $w$, and 
$u$ is the sequence of utilities defined by $u_t = \sum_{i=0}^{t} \mu \cdot M^i \cdot w^\transpose$
for all $t\geq 0$. With this definition in mind, we also denote the optimal expected 
value of a Markov chain~$M$ by $\val(u,T)$.
The best-case expected value, defined using $\sup$ instead of $\inf$
in the above definition, can be computed as the opposite of the worst-case expected value for
the Markov chain with all weights multiplied by~$-1$.

\medskip\noindent{\bf Exact value problem with expected stopping time.}
Given a Markov chain $\tuple{M,\mu,w}$, a rational stopping time $T$, and a rational 
threshold $\theta$, decide whether the optimal expected value of $M$
with expected stopping time $T$ is below $\theta$, i.e., whether $\val(M,T) < \theta$.

\medskip\noindent{\bf Approximation of the value with expected stopping time.}
We also consider an approximate version of the exact value problem, where the
goal is to compute, given $\epsilon > 0$, a value $v_{\epsilon}$ such that
$\abs{\val(M,T) - v_{\varepsilon}} \leq \epsilon$. We say that $v_{\epsilon}$
is an \emph{approximation with additive error $\epsilon$} of the optimal value.

\subsection{Hardness of the exact value problem}\label{sec:exact}

This section is devoted to the proof of the following result, which establishes
the inter-reducibility of the exact value problem, the Positivity problem, 
and the Markov Reachability problem (defined in Section~\ref{sec:reverse-Skolem}).

\begin{thm}\label{thm:exact-hard}
The Positivity problem, the inequality variant of the Markov Reachability problem,
and the exact value problem with expected stopping time are inter-reducible.
\end{thm}

The decidability status of the Positivity problem is a longstanding open question, 
although decidability is known for dimension $n\leq 5$~\cite[Section~4]{OW14}. Therefore,
constructing an algorithm to compute the exact value of a Markov chain with 
expected stopping time $T$ would require the significant advances in number theory
that are necessary to solve the Positivity problem~\cite[Section~5]{OW14}.

We also show the converse reduction from the exact value problem to the Positivity problem.
Hence proving the undecidability of the exact value problem would also be a major
breakthrough, as it would entail the undecidability of the Positivity problem.

The proof of Theorem~\ref{thm:exact-hard} is presented in the rest of this section.

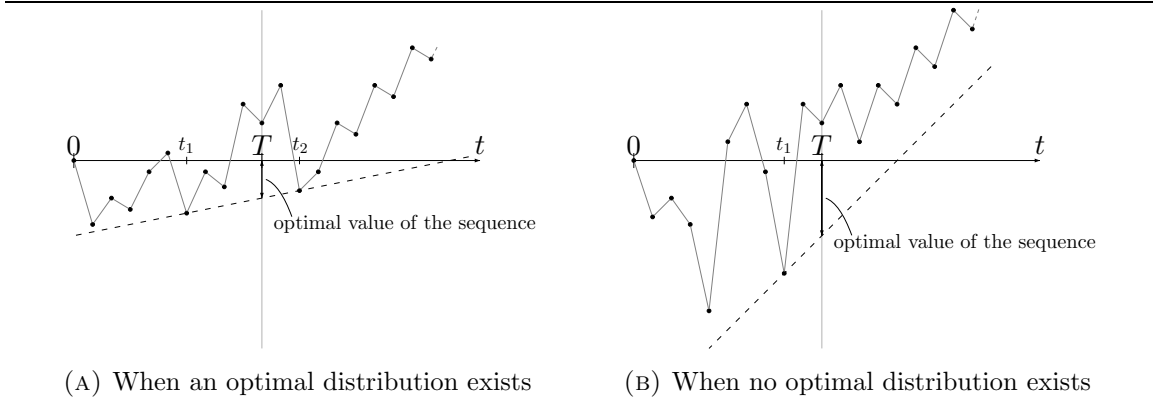
\begin{figure}[!tbp]
  \centering
  \hrule
  \begin{subfigure}[b]{.48\linewidth}
     \input{figures/geometry-intuition-1.tex}
     \caption{When an optimal distribution exists}\label{fig:geometry1}
  \end{subfigure}~
  \begin{subfigure}[b]{.48\linewidth}
     \input{figures/geometry-intuition-2.tex}
     \caption{When no optimal distribution exists\label{fig:geometry2}}
  \end{subfigure}
  \hrule
  \caption{Geometric interpretation of the value of a sequence of utilities.}\label{fig:geom}
\end{figure}

\subsubsection{Geometric interpretation}\label{sec:Geometric-Interpretation}

A geometric interpretation for (arbitrary) sequences of real numbers and expected stopping-time 
was developed in previous work~\cite{CD19}. We recall the main result
in this section. The rest of our technical results is independent from~\cite{CD19} 
(see also Comparison with related work in Section~\ref{sec:introduction}).

It is known that bi-Dirac distributions are sufficient for optimal expected value,
namely for all sequences $u = u_0, u_1, \dots$ of utilities, for all time bounds~$T$, the following holds~\cite{CD19}:
\begin{align*}
&\inf \{\E_{\delta}(u) \mid \delta \in \Delta \land \E_{\delta} = T\} = \\
&\inf \{\E_{\delta}(u) \mid \delta \in \Delta^{\upuparrows} \land \E_{\delta} = T\}.
\end{align*}
Moreover the value of the expected utility of the sequence $u$ 
under a bi-Dirac distribution with support $\{t_1,t_2\}$ (where $t_1 < T < t_2$)
and expected time $T$ is given by
\begin{equation}
   u_{t_1} + \frac{T - t_1}{t_2 - t_1} \cdot (u_{t_2} - u_{t_1}). \label{eqn:val-bi-Dirac}
\end{equation}
As illustrated in \figurename~\ref{fig:geometry1}, this value is obtained as the
intersection of the vertical axis at $T$ and the line that connects
the two points $(t_1, u_{t_1})$ and $(t_2, u_{t_2})$. Intuitively,
the optimal value of a sequence of utilities is obtained by choosing the two points $t_1$ and $t_2$
such that the connecting line intersects the vertical axis at $T$
as low as possible.

It is always possible to fix a value of $t_1$ such that it is sufficient 
to consider bi-Dirac distributions with support containing $t_1$ to compute
the optimal value (because $t_1 \leq T$ is to be chosen among a finite set 
of points), but the optimal value of $t_2$ may not exist, as in \figurename~\ref{fig:geometry2}. 
In that case, the value of the sequence of utilities is obtained as $t_2 \to \infty$. 

Given such a value of $t_1$, let $\nu = \inf_{t_2 \geq T}  \frac{u_{t_2} - u_{t_1}}{t_2 - t_1}$, 
and Lemma~\ref{lem:geometric-interpretation} shows 
that $ u_t \geq f_u(t)$, for all $t \geq 0$ where $f_u(t) = u_{t_1} + (t - t_1) \cdot \nu.$
The optimal expected utility is
\begin{align*}
\val(u,T) & = \min_{0 \leq t_1 \leq T} \,\, \inf_{t_2 \geq T} \,\, u_{t_1} + \frac{T - t_1}{t_2 - t_1} \cdot (u_{t_2} - u_{t_1}) \\
& = \min_{0 \leq t_1 \leq T} \,\, u_{t_1} + (T - t_1) \cdot \nu \\
& = f_u(T),
\end{align*}
hence $f_u(T)$ is the optimal value.

\begin{lem}[Geometric interpretation~\cite{CD19}]\label{lem:geometric-interpretation}
For all sequences $u$ of utilities:

\begin{itemize} 
\item if $u_t \geq a \cdot t + b$ for all $t \geq 0$,
then the optimal value of the sequence $u$ is at least $a \cdot T + b$;

\item we have $u_t \geq f_u(t)$ for all $t \geq 0$, 
and the optimal expected value of $u$ is $f_u(T)$.
\end{itemize} 
\end{lem}

It follows from Lemma~\ref{lem:geometric-interpretation} that the optimal value of 
the sequence $u$ is the largest possible value at $T$ of a line that lies
below $u$: $\val(u,T) = \sup \{f(T) \mid \exists a,b \cdot \forall t: f(t) = a\cdot t + b \leq u_t\}$.

\subsubsection{Reduction of the Positivity problem to the exact value problem}\label{sec:Skolem}

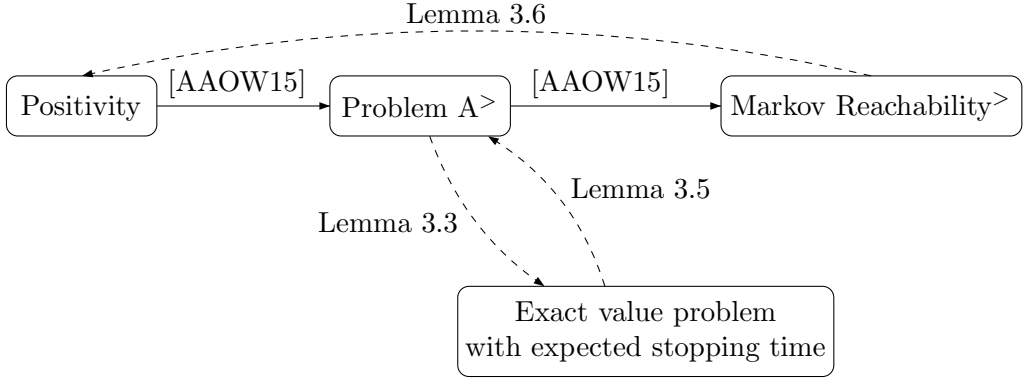
\begin{figure}[!tb]
  \begin{center}
    \hrule
      \input{figures/reduction-chains-compact.tex}
    \hrule
      \caption{Known reductions (solid lines), and reductions established in this paper (dashed lines). \label{fig:reduction-chains}}
  \end{center}
\end{figure}

It is known that the Positivity problem can be reduced to the inequality variant
of~\cite[Problem~A]{AAOW15}, defined below as~A${}^{>}$. A subsequent reduction
of~A${}^{>}$ to the exact value problem with expected stopping time establishes 
one direction of Theorem~\ref{thm:exact-hard}. We present such a reduction in the
proof of Lemma~\ref{lem:pos-to-exact} (see also \figurename~\ref{fig:reduction-chains}).

\medskip\noindent{\bf Problem~A${}^{=}$~\cite{AAOW15}.}
Given a $n \times n$ aperiodic\footnote{Although in the original 
formulation of Problem~A, the stochastic matrix $M$
need not be aperiodic, the reduction of the Positivity problem
to Problem~A produces stochastic matrices that define 
aperiodic Markov chains (even ergodic unichains)~\cite{AAOW15}.} stochastic matrix $M$ with rational entries, 
an initial distribution $\mu = (1,0,\dots,0)$, and a vector $z \in \{0,1,2\}^n$, decide whether 
there exists an integer $t \geq 1$ such that $\mu \cdot M^t \cdot z^\transpose = 1$.

\medskip\noindent{\bf Problem~A${}^{>}$~\cite{AAOW15}.}
Given a $n \times n$ aperiodic stochastic matrix $M$ with rational entries, an initial 
distribution $\mu = (1,0,\dots,0)$, and a vector $z \in \{0,1,2\}^n$, decide whether 
there exists an integer $t \geq 1$ such that $\mu \cdot M^t \cdot z^\transpose > 1$.
\medskip

Problems~A${}^{=}$ and A${}^{>}$ are difficult to solve only in the case where $\mu \cdot M^t \cdot z^\transpose$
converges to~$1$ as $t \to \infty$. Otherwise, an argument based on the
definition of convergence to a limit shows that the problems are decidable~\cite[Theorem~1]{KA04}.
Note that $\lim_{t \to \infty} \mu \cdot M^t$ exists since $M$
is aperiodic, and the limit is the steady-state vector $\pi$, which is algorithmically computable.
Hence we can assume that the instances of Problem~A${}^{>}$ are such that 
\begin{equation}
 \pi \cdot z^\transpose = 1. \label{eqn:pi-z}
\end{equation}

Moreover, without loss of generality, we can modify $M$ such 
that there is no incoming transition to the initial vertex~$1$ (remember that $\mu(1) = 1$)
by creating a copy of the initial vertex, and redirecting the transitions to $1$
towards the copy vertex. Thus we require the matrix $M$ in Problem~A${}^{>}$ to define
a Markov chain consisting of an initial vertex $1$ with no incoming transition.
This may however increase the dimension of the matrix by $1$.

\begin{lem}\label{lem:pos-to-exact}
Problem~A${}^{>}$ can be reduced to the exact value problem with expected stopping time.
\end{lem}

\begin{cor}\label{cor:pos-to-exact}
The Positivity problem can be reduced to the exact value problem with expected stopping time.
\end{cor}

The proof of Lemma~\ref{lem:pos-to-exact} is organized as follows: 
we first recall basic results from the theory of Markov chains, then
present a reduction of    
Problem~A${}^{>}$ to the exact value problem with expected stopping time,
and establish its correctness.

\smallskip\noindent\emph{Basic results.} First we show that, given an aperiodic 
Markov chain $\tuple{M,\mu,w}$ that has a single recurrent class,
there exist vectors $x,y$ such that the expected utility after $t$ steps tends to 
$\mu \cdot x^\transpose \cdot t + \mu \cdot y^\transpose$ as $t \to \infty$, formally:
\begin{equation} \lim_{t \to \infty} \left\lvert \sum_{i=0}^{t-1} \mu \cdot M^i \cdot w^\transpose - \mu \cdot (x^\transpose \cdot t + y^\transpose)\right\rvert = 0. \label{eqn:convergence}
\end{equation}

The vector $x$ is called the \emph{gain} per time unit, and $y$ is the \emph{relative-gain} vector.
They can be computed by solving the following equations (following~\cite[Section~4.5]{Gallager}):
\begin{align*}
\left\{ \begin{array}{l}
  x_i = \pi \cdot w^\transpose \text{ for all vertices } i \in V \\[+2pt]
  y^\transpose = M \cdot (y-x)^\transpose + w^\transpose \numberthis \label{eqn:y} \\[+2pt]
  \pi \cdot y^\transpose = 0  \\[+2pt]
\end{array} \right.
\end{align*}
\indent The number $g = \pi \cdot w^\transpose$ is the gain per time unit. 
Note that $x = g \cdot e$ where $e = (1,1,\dots,1)$,
and that $M \cdot x^\transpose = x^\transpose = g \cdot e^\transpose$ because $M$ 
is stochastic and the sum of the elements in each of its row is $1$.
It follows that Equation~\eqref{eqn:y} can be written as 
$y^\transpose = M \cdot y^\transpose + w^\transpose - x^\transpose$,
and by $t-1$ successive substitutions of $y^\transpose$, we get 
\begin{align*}
 y^\transpose & = M^t \cdot y^\transpose + \sum_{i=0}^{t-1} M^i \cdot w^\transpose - \sum_{i=0}^{t-1} M^i \cdot x^\transpose \\
              & = M^t \cdot y^\transpose + \sum_{i=0}^{t-1} M^i \cdot w^\transpose - t \cdot x^\transpose \\
\end{align*}
\indent Then, the rate of convergence of the expected utility evaluates as follows, for all $t \geq 1$:
\begin{align*}
  & \sum_{i=0}^{t-1} \mu \cdot M^i \cdot w^\transpose - \mu \cdot (x^\transpose \cdot t + y^\transpose) \\
= & \sum_{i=0}^{t-1} \mu \cdot M^i \cdot w^\transpose - \mu \cdot x^\transpose \cdot t \\
  & - \mu \cdot M^t \cdot y^\transpose - \mu \cdot \sum_{i=0}^{t-1} M^i \cdot w^\transpose + t \cdot \mu \cdot x^\transpose    \\
= & - \mu \cdot M^t \cdot y^\transpose \numberthis \label{eqn:utility-convergence}
\end{align*}
which tends to $-\pi \cdot y^\transpose = 0$ as $t \to \infty$, establishing~\eqref{eqn:convergence}.

In the case of an aperiodic Markov chain with multiple recurrent classes,
the gain and relative gain satisfying Equation~\eqref{eqn:convergence} 
can be computed as the linear combination
of the vectors $x,y$ obtained for each recurrent class, where the coefficient
in the linear combination is the mass of probability that reaches (in the limit) the 
recurrent class from the initial distribution~$\mu$.

\smallskip\noindent\emph{Reduction.} The reduction from Problem~A${}^{>}$
to the exact value problem with expected stopping time is as follows.
Given an instance $(M,\mu,z)$ of Problem~A${}^{>}$, we construct an instance 
of the exact value problem in two stages. First, 
let $w^\transpose = z^\transpose - M \cdot z^\transpose$
be a reward vector defining a Markov chain $\tuple{M,\mu,w}$.
We explain later why $w$ is defined in this way. 

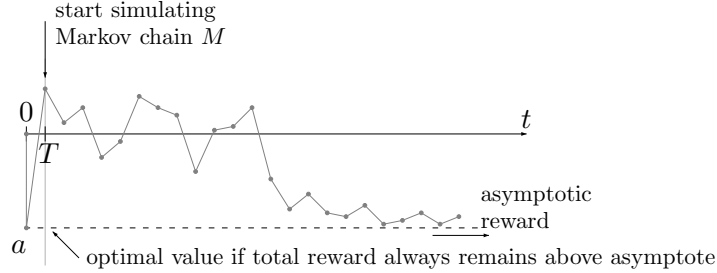
\begin{figure}[!tb]
  \begin{center}
    \hrule
      \input{figures/Positivity-reduction.tex}
    \hrule
      \caption{Reduction of the Positivity problem to the exact value problem. \label{fig:Positivity-reduction}}
  \end{center}
\end{figure}

We proceed to the second stage of the construction, and define the instance
of the exact value problem, namely the Markov chain
$\tuple{M',\mu',w'}$, the expected time $T$, and the threshold~$\theta$.
The key idea of the construction is illustrated in \figurename~\ref{fig:Positivity-reduction}.
Given the Markov chain $\tuple{M,\mu,w}$, we can compute its asymptotic expected
utility, shown as the dashed line $\mu \cdot (x^\transpose \cdot t + y^\transpose)$ 
in \figurename~\ref{fig:Positivity-reduction} which also plots the sequence $u_{t-1}$ for
$t \geq 1$. Note that by Equation~\eqref{eqn:convergence}
we have $\lim_{t \to \infty} \left\lvert u_{t-1} - \mu \cdot (x^\transpose \cdot t + y^\transpose)\right\rvert = 0$ and by Equations~\eqref{eqn:y} $x = \pi \cdot z^\transpose - \pi \cdot M \cdot z^\transpose = 0$.
 
We construct an instance of the exact value problem in such a way that, if the utility of $M$
always remains above its asymptote, then the optimal value is the value
of the asymptote at time $T$, and otherwise, the optimal value is strictly smaller.
We achieve this by having an initial vertex with weight $a$ such that,
if the Markov chain $\tuple{M,\mu,w}$ is executed (simulated) after the initial vertex, 
then the weight $a$ lies exactly on the asymptote of $\tuple{M,\mu,w}$ (see \figurename~\ref{fig:Positivity-reduction}
and the geometric interpretation in Section~\ref{sec:Geometric-Interpretation}). 
Since we simulate $\tuple{M,\mu,w}$ after one time step, the value of $a$ is chosen such that
the point $(0,a)$ belongs to the line $\mu \cdot (x^\transpose \cdot t + y^\transpose)$.
Since $\mu = (1,0,\dots,0)$ in Problem~A${}^{>}$, we have $a = y(1)$.
To recover the original behavior of the Markov chain~$\tuple{M,\mu,w}$, we subtract $a$ from the weight of 
the initial vertex of $M$, thus $w'(1) = w(1) - a$. 
As we assumed that the initial vertex in $M$ has no incoming transition, 
it is never re-visited later.
We take $T=1$ and the value of the asymptote 
at time $T$ is $\mu \cdot (x^\transpose + y^\transpose) = x(1) + y(1) = a$,
which we define as the threshold $\theta$ of the exact value problem, thus $\theta = a$.

Formally, the instance of the exact value problem is defined as follows:
$$ w' = \left(\begin{array}{c} a \\ w(1) - a \\ w(2) \\  \vdots \\ w(n)  \end{array} \right),
\quad 
M' = \left(\begin{array}{cc} 0 & \mu \\ 0 & M \end{array} \right),
\begin{array}{lll}
\mu' = (1,0,\dots,0), \\
T = 1, \\
\theta = a \\
\end{array}
$$
where $a = y(1)$ and $y$ is the relative-gain vector of the Markov chain $\tuple{M,\mu,w}$.
Note that the initial vertex of $M$ has no incoming transition (in $M$),
and thus the sequence of expected utilities in $M'$ indeed simulates the sequence of
expected utilities in $M$, and the asymptotic expected utilities as well as the 
steady-state vectors of $\tuple{M,\mu,w}$ and $\tuple{M',\mu',w'}$ coincide.

\smallskip\noindent\emph{Correctness of the reduction.} To establish the correctness of the reduction, 
we show the following equivalences: 
\begin{enumerate}
\item the optimal expected value of $M'$ with expected stopping time $T$ is smaller than $\theta$
(i.e., the answer to the exact value problem is {\sc Yes})
if and only if the utility sequence of $M$ eventually drops below its asymptote;

\item the utility sequence of $M$ eventually drops below its asymptote if and 
only if $\mu \cdot M^t \cdot z^\transpose > 1$ for some $t \geq 1$ (i.e., the answer to Problem~A${}^{>}$ is {\sc Yes}).
\end{enumerate}

To show the first equivalence, consider the first direction and assume that 
the value of $M'$ is smaller than $\theta$. Given that the line $\mu \cdot (x^\transpose \cdot t + y^\transpose)$
has value $\theta$ at $t=T$, it follows from Lemma~\ref{lem:geometric-interpretation}
that the utility sequence of $M'$ does not always remain above that line, 
and thus the utility sequence of $M$ eventually drops below its asymptote.

Now consider the second direction of the first equivalence and assume that the
utility sequence of $M$ eventually drops below its asymptote. Then the 
utility sequence of $M'$ drops below the line $\mu \cdot (x^\transpose \cdot t + y^\transpose)$,
say at time $t_2 \geq 1$. 
We construct a distribution $\delta$ with $\E_{\delta} = T$ such that the value
of the expected reward under $\delta$ is less than $\mu \cdot (x^\transpose \cdot T + y^\transpose) = \theta$
(which implies that the optimal value, obtained as the infimum over all distributions, is also below $\theta$).

We consider two cases: $(1)$ if $t_2 = 1$ (i.e., $t_2 = T$), consider the distribution $\delta$ such that 
$\delta(t_2) = 1$ (note that $\E_{\delta} = T$) and the result follows immediately; $(2)$ 
otherwise, $t_2 > 1$ and consider the bi-Dirac distribution with support $\{t_1,t_2\}$ where $t_1 = 0$. 
Note that $t_1 < T < t_2$ and the value of the expected reward under this distribution is given by the value 
at time $T$ of the line connecting the point $(t_1,a)$ and a point below 
the asymptote (at $t_2$), see Equation~\eqref{eqn:val-bi-Dirac}.
This value is below the value $\theta$ of the asymptote at time $T$ since $(t_1,a)$ 
is on the asymptote, and the other point (at $t_2$) 
is strictly below the asymptote. 

To show the second equivalence, note that by Equation~\eqref{eqn:utility-convergence}
the utility sequence of $M$ eventually drops below its asymptote
if and only if $- \mu \cdot M^t \cdot y^\transpose < 0$ for some $t \geq 1$.
Hence we can establish the second equivalence by showing that 
$- \mu \cdot M^t \cdot y^\transpose < 0$ if and only if $\mu \cdot M^t \cdot z^\transpose > 1$.
This is where the value of $w$ is important. The result holds if $y = z - e$,
and we just need to show that $y=z-e$ satisfies Equations~\eqref{eqn:y}, namely that 
\begin{align*}
 & (z - e)^\transpose = M \cdot (z - e - x)^\transpose + w^\transpose   \\
 & \pi \cdot (z - e)^\transpose = 0  
\end{align*}
that is
\begin{align*}
 & (z - e)^\transpose = M \cdot z^\transpose - e^\transpose - x^\transpose + z^\transpose - M \cdot z^\transpose    \\
 & \pi \cdot z^\transpose - \pi \cdot e^\transpose= 0  
\end{align*}
which hold since $x = 0$ and $\pi \cdot z^\transpose = 1 = \pi \cdot e^\transpose$ (Equation~\eqref{eqn:pi-z}). This concludes the proof of Lemma~\ref{lem:pos-to-exact}.

Using the reduction of the Positivity problem to Problem~A${}^{>}$~\cite{AAOW15},
we obtain Corollary~\ref{cor:pos-to-exact}, showing that a decidability result for the exact value problem 
would imply the decidability of the Positivity problem, which is a longstanding open question.

\subsubsection{Reduction of the exact value problem to the Positivity problem}\label{sec:reverse-Skolem}

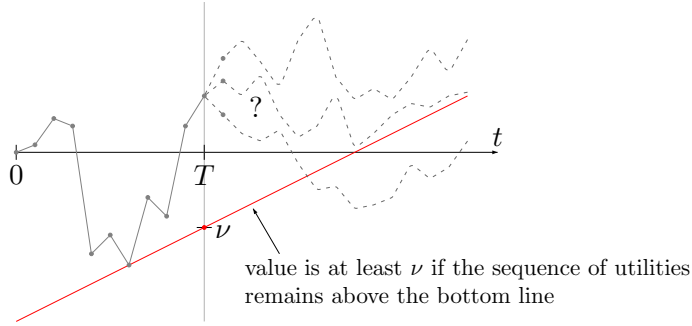
\begin{figure}[!tb]
  \begin{center}
    \hrule
      \input{figures/Positivity-oracle.tex}
    \hrule
      \caption{Solving the exact value problem using an oracle for Problem~A${}^{>}$. \label{fig:Positivity-oracle}}
  \end{center}
\end{figure}

We present the converse reduction of Section~\ref{sec:Skolem}, showing that to potentially prove
the exact value problem is undecidable would require such a proof for the Positivity problem as well.
We sketch the reduction by showing how the exact value problem can be solved using 
an oracle for Problem~A${}^{>}$, illustrated in \figurename~\ref{fig:Positivity-oracle}, and then
present a reduction of Problem~A${}^{>}$ to the Positivity problem.

\begin{lem}\label{lem:exact-to-A}
The exact value problem with expected stopping time can be reduced to Problem~A${}^{>}$. 
\end{lem}

\begin{proof}[Proof]
Given a Markov chain $\tuple{M,\mu,w}$ with expected stopping time $T$ and threshold $\theta$, 
we solve the exact value problem using 
an oracle for Problem~A${}^{>}$ as follows. 
First, if $u_T < \theta$ then the answer to the exact value problem is {\sc Yes}.
Otherwise, we compute the value of utilities $u_t = \sum_{i=0}^{t} \mu \cdot M^i \cdot w^\transpose$
for all $0 \leq t < T$, and let $b = \max_{0 \leq t < T} \frac{u_t-\theta}{t-T}$.
Consider the \emph{bottom line} of equation $b \cdot (t-T) + \theta$ and observe
that $u_t \geq b \cdot (t-T) + \theta$ for all $0 \leq t \leq T$ (see \figurename~\ref{fig:Positivity-oracle}).
By the geometric interpretation lemma (Lemma~\ref{lem:geometric-interpretation}), 
it suffices to determine whether the sequence of utilities ever drops below the bottom line
to answer the exact value problem.

For simplicity of presentation, we assume that the recurrent classes of the Markov chain are all aperiodic.
The case of periodic recurrent classes (say $C_1, C_2, \dots, C_l$ with respective period $d_1, \dots, d_l$) can be solved analogously by applying the
procedure to each initial distribution $\mu, \mu \cdot M, \mu \cdot M^2, \dots, \mu \cdot M^{d-1}$,
and the transition matrix $M^d$ where $d$ is the period of the Markov chain, i.e., $d=\lcm \{d_1, \dots, d_l \}$.

Using Equations~\eqref{eqn:y}, we can compute the \emph{asymptote} of the sequence of utilities
as $\mu \cdot x^\transpose \cdot (t+1) + \mu \cdot y^\transpose$. 
Comparing the slope of the bottom line and the slope of the asymptote, we have the following cases:

\begin{itemize}
\item if $b < \mu \cdot x^\transpose$, then from some point on the utilities
always remain above the bottom line. 
Such a point can be computed using the convergence rate of Markov chains (Appendix~\ref{sec:rate-convergence}).
Then the answer to the exact value problem is {\sc No} if $u_t \geq b \cdot (t-T) + \theta$ 
for all (the finitely many) values of $t$ up to that point. 
Otherwise the answer is {\sc Yes}.

\item if $b > \mu \cdot x^\transpose$, then eventually the utility gets below
the bottom line, thus the answer to the exact value problem is {\sc Yes}.

\item if $b = \mu \cdot x^\transpose$, then either the bottom line is different
from the asymptote, and we can reuse one of the above cases, or the bottom 
line is equal to the asymptote, and the condition for the sequence of utilities 
to eventually drop below the bottom line is that $- \mu \cdot M^t \cdot y^\transpose < 0$ for some $t \geq 1$
(Equation~\eqref{eqn:utility-convergence}),
which is equivalent to $\mu \cdot M^t \cdot (e+y)^\transpose > 1$ for some $t \geq 1$ (where $e = (1,1,\dots,1)$). 
We cannot immediately use an oracle for Problem~A${}^{>}$ with $z = e+y$ because
Problem~A${}^{>}$ requires $z \in \{0,1,2\}^n$. However, we can obtain a valid 
instance of Problem~A${}^{>}$ as follows. First we apply a scaling
factor to $y$ to ensure that its components are in the interval $[-1,1]$. 
Scaling the vector $y$ does not affect the answer to the original question 
(whether $- \mu \cdot M^t \cdot y^\transpose < 0$). Then we construct a Markov
chain $\tuple{M',\mu',w'}$ as follows: for each vertex $v \in V$ of the Markov
chain $M$, we create a copy $v^0$ and $w'(v^0) = 0$.
Define $w'(v) = 1$ if $y(v) \geq 0$ and $w'(v) = -1$ if $y(v) < 0$, where $y(v)$ 
is the component in $y$ corresponding to vertex $v$.
For each transition $v \xrightarrow{p} u$ (i.e., $M_{v,u} = p$),
we have the transitions $v \xrightarrow{p \cdot \abs{y(u)}} u$ and $v \xrightarrow{p \cdot (1-\abs{y(u)})} u^0$
in $M'$.
The outgoing transitions from the copy $v^0$ are the same as from $v$. 
Note that the weights in $w'$ are in the set $\{-1,0,1\}$, and therefore
the vector $z = e+w'$ is in $\{0,1,2\}^{2n}$.
We now call the oracle for Problem~A${}^{>}$ with the Markov chain $M'$ and $z = e+w'$
which gives the answer to the exact value problem (in this way we transferred
the value of $y$ into the transition probabilities of $M'$, and note that the dimension 
of $M'$ is twice the dimension of $M$).  \qedhere 
\end{itemize}
\end{proof}

To obtain the inter-reducibility result of Theorem~\ref{thm:exact-hard}, 
we need to show that Problem~A${}^{>}$ can be reduced to the Positivity problem,
which we establish by showing that the inequality version of the Markov reachability problem (defined below)
can be reduced to the Positivity problem, as it is known that Problem~A${}^{>}$ can be reduced to the 
inequality variant of the Markov reachability problem~\cite{AAOW15} (see also \figurename~\ref{fig:reduction-chains}). 
This is a straightforward result established in Lemma~\ref{lem:markov-to-pos}.

\medskip\noindent{\bf Markov reachability${}^{>}$ problem~\cite{AAOW15}.}
Given a square stochastic matrix $M$ with rational entries and a rational number $r > 0$, 
decide whether there exists an integer $t \geq 1$ such that $M^t_{1,2} > r$. 

\medskip\noindent{\bf Markov reachability${}^{=}$ problem~\cite{AAOW15}.}
Given a square stochastic matrix $M$ with rational entries and a rational number $r > 0$, 
decide whether there exists an integer $t \geq 1$ such that $M^t_{1,2} = r$.

\begin{lem}\label{lem:markov-to-pos}
The Markov reachability${}^{>}$ problem can be reduced to the Positivity problem. 
\end{lem}

\begin{proof}
The reduction of the Markov reachability${}^{>}$ problem
to the Positivity problem is as follows. Given a $n \times n$ stochastic matrix $M$
and a rational number $r > 0$, define the $(n+3)\times(n+3)$ matrix $P$ as follows,
where $0_{n\times1}$ is the zero column-vector of length $n$, and
$e_1 = (1,0,0,\dots,0)$ and $e_2 = (0,1,0,\dots,0)$ (so that $e_1 \cdot M \cdot e_2^\transpose = M_{1,2}$):

$$ P = \left(\begin{array}{cccc} M     & 0_{n \times 1} & 0_{n \times 1} & e_2^\transpose \\
                                 e_1 \cdot M & 0        & -r             & 0 \\
                                 0_{1 \times n}     & 0              & 1              & 1 \\
                                 0_{1 \times n}     & 0              & 0              & 0 \end{array}  \right) $$
All entries of $P$ are rational numbers, hence there exists $d \in \nat$ such that $\widetilde{P} = d \cdot P$
is an integer matrix.
It is easy to show by induction that for all $t \geq 2$:
$$ \widetilde{P}^t = d^t \cdot \left(\begin{array}{cccc} M^t & 0_{n \times 1} & 0_{n \times 1} & M^{t-1} \cdot e_2^\transpose \\
                                               e_1 \cdot M^{t} & 0    & -r             & \mathbf{e_1 \cdot M^{t-1} \cdot e_2^\transpose - r}\\
                                               0_{1 \times n}     & 0              & 1              & 1 \\
                                               0_{1 \times n}     & 0              & 0              & 0 \end{array}  \right) $$
It follows that $\widetilde{P}^t_{n+1,n+3} = d^t \cdot (e_1 \cdot M^{t-1} \cdot e_2^\transpose - r) = d^t \cdot (M^{t-1}_{1,2} - r)$,
and therefore $\widetilde{P}^t_{n+1,n+3} > 0$ if and only if $M^{t-1}_{1,2} > r$ (for all $t \geq 2$). 
Since $\widetilde{P}_{n+1,n+3} = 0$, we conclude that 
there exists an integer $t \geq 1$ such that $\widetilde{P}^t_{n+1,n+3} > 0$
if and only if there exists an integer $t \geq 1$ such that $M^t_{1,2} > r$,
which concludes the reduction by rearranging the order of the rows and columns
of $\widetilde{P}$, and the desired result follows. 
\end{proof}

The results of Lemma~\ref{lem:pos-to-exact},~\ref{lem:exact-to-A}~and~\ref{lem:markov-to-pos}
establish Theorem~\ref{thm:exact-hard}. 
The reduction in Lemma~\ref{lem:markov-to-pos} can easily be adapted to show that 
the Markov reachability${}^{=}$  problem can be reduced to the Skolem problem and 
thus these problems are inter-reducible with Problem~A${}^{=}$ (the adaptation is to replace 
$\widetilde{P}_{n+1,n+3}$ by an arbitrary nonzero value, which has no effect
on the value of the powers of $\widetilde{P}$).

\begin{thm}\label{thm:exact-pos}
The Skolem problem, Problem~A${}^{=}$, and the Markov reachability${}^{=}$ problem 
are inter-reducible.
\end{thm}

\subsection{Approximation of the optimal value}\label{sec:approx}

We can compute an approximation of the optimal value with additive error by considering 
an approximation $u'$ of the exact sequence $u$ of expected utilities 
of the Markov chain as follows: for a large number of time steps, 
let the approximate sequence $u'$ be equal to $u$, and then
from some point on it switches to the value of the limit (asymptotic, and possibly periodic) 
sequence of expected utilities at the steady-state distribution(s).
By taking the switching point large enough, the approximation sequence~$u'$ 
can be made arbitrarily close to the exact sequence~$u$.
We show that the value of the sequences $u'$ approximates 
arbitrarily closely the (exact) optimal value of $u$.

By the results of Section~\ref{sec:Geometric-Interpretation},
the optimal expected value of any sequence $u'$ of utilities 
is given by the expression

\begin{equation}
\val(u',T) = \min_{0 \leq t_1 \leq T} \,\, \inf_{t_2 \geq T} \,\,
\frac{u'_{t_1} (t_2 - T) + u'_{t_2} (T - t_1)}{t_2 - t_1}. \label{eqn:optimal}
\end{equation}

We can effectively compute the value of $\val(u',T)$ when $u'$ is an ultimately
periodic sequence, i.e. $u' = A.C^{\omega}$ where $A, C$ are finite sequences (with $C$ nonempty):
we show in Lemma~\ref{lem:val-cycle} that the infinite range of $t_2$ in the 
expression~\eqref{eqn:optimal} can be replaced by a finite range, because the 
optimal value is obtained either by taking $t_2$ before the first repetition of the cycle~$C$,
or by taking $t_2 \to \infty$ (i.e., if repeating the cycle once improves the value,
then repeating the cycle infinitely often improves the value even more). 
Let $S_A$ and $S_C$ be the sum of the weights in $A$ and $C$ respectively, 
let $M_C = \frac{S_C}{\abs{C}}$ be the average weight of the cycle~$C$. 

\begin{lem}\label{lem:val-cycle}
The optimal value of an ultimately periodic sequence $u = A.C^{\omega}$ is 
$\val(u,T) = \min\{E_1,E_2\}$ where
\begin{align*}
E_1 & = \min_{0 \leq t_1 \leq T} \,\, \min_{T \leq t_2 \leq \abs{A} + \abs{C}} \,\, 
\frac{u_{t_1} (t_2 - T) + u_{t_2} (T - t_1)}{t_2 - t_1}, \text{and} \\[+6pt]
E_2 & = \min_{0 \leq t_1 \leq T} u_{t_1} + M_C \cdot (T-t_1). 
\end{align*}
If $T \geq \abs{A} + \abs{C}$, then $\val(u,T) = \min_{0 \leq t_1 \leq \abs{A} + \abs{C}} u_{t_1} + M_C \cdot (T-t_1)$. 
\end{lem}

\begin{proof}
The expression $E_1$ is the expression~\eqref{eqn:optimal} where the range of
$t_2$ is the interval $[T, \abs{A} + \abs{C}]$. We now show that the expression
$E_2$ corresponds to $t_2 \geq \abs{A}$ (assuming always $t_2 \geq T$),
which covers  $t_2 \geq \abs{A} + \abs{C}$. 
For all $t_2 \geq \abs{A}$, we can express $t_2$ as 
$t_0 + k \cdot \abs{C}$ where $\abs{A} \leq t_0 \leq \abs{A} + \abs{C}$ 
and $k \geq 0$.
We have two cases:
\begin{itemize}

\item If $T \leq \abs{A} + \abs{C}$, then the expression~\eqref{eqn:optimal} gives $\val(u,T) =$
$$
\min_{0 \leq t_1 \leq T} \,\, \min_{\max(T,\abs{A}) \leq t_0 \leq \abs{A} + \abs{C}} \,\, \inf_{k \geq 0} \,\, \frac{u_{t_1} (t_0 + k \cdot \abs{C} - T) + (u_{t_0} + k \cdot S_C) (T - t_1)}{t_0 + k \cdot \abs{C} - t_1}
$$
where the numerator and denominator of the fraction are linear in $k$. 
Such functions $\frac{a\cdot k+b}{c \cdot k + d}$ are monotone (their first derivative
has constant sign), and note that the denominator $t_0 + k \cdot \abs{C} - t_1$ is nonzero for all $k \geq 0$.
It follows that the infimum over $k \geq 0$ is obtained either for $k=0$, which 
is covered by the expression $E_1$, or for $k \to \infty$, which gives
the expression $\frac{u_{t_1} \cdot \abs{C} + S_C \cdot (T- t_1)}{\abs{C}} = u_{t_1} + M_C \cdot (T-t_1)$,
corresponding to $E_2$.

\item If $T \geq \abs{A} + \abs{C}$, then the same reasoning as above shows that 
$\val(u,T) = \min_{0 \leq t_1 \leq T} u_{t_1} + M_C \cdot (T-t_1)$, as the 
range of $t_2$ in expression $E_1$ is empty. 
Now observe that if $t_1 = t_0 + k \cdot \abs{C}$ where $\abs{A} \leq t_0 \leq \abs{A} + \abs{C}$, then 
$u_{t_1} + M_C \cdot (T-t_1) = u_{t_0} + M_C \cdot (T-t_0) + k \cdot S_C - M_C \cdot k \cdot \abs{C} = u_{t_0} + M_C \cdot (T-t_0)$
and thus we can restrict the range of $t_1$ to $[0, \abs{A} + \abs{C}]$. The
result follows. \qedhere
\end{itemize}
\end{proof}

We show that for a sequence $u'$ of utilities that approximates the sequence $u$,
the value of $u'$ approximates the value of $u$ and the error can be bounded. 
Precisely, if the weights in a Markov chain are shifted by at most $\eta$, then
the optimal expected value of the Markov chain with expected stopping time $T$ 
is shifted by at most $\eta \cdot (T+1)$.
Consider $w'$ such that $\abs{w'(v) - w(v)} \leq \eta$ for all vertices $v \in V$,
and consider the sequences $u$ and $u'$ of utilities of a path according to $w$ and 
$w'$ respectively. Then we have $\abs{u'_t - u_t} \leq (t+1) \cdot \eta$ for all $t \geq 0$,
and for all distributions~$\delta$ with $\E_{\delta} = T$:
\begin{align*} 
\Abs{\sum_i \delta(i) \cdot u'_{i} - \sum_i \delta(i) \cdot  u_{i}} & \leq \sum_i \delta(i) \cdot \abs{u'_{i} - u_{i}}\\
 & \leq \sum_i \delta(i) \cdot (i+1) \cdot \eta\\
 & = (T+1) \cdot \eta.
\end{align*}
It follows that $\abs{\val(u',T) - \val(u,T)} \leq (T+1) \cdot \eta$, 
that is the value of the sequence is shifted by at most $(T+1) \cdot \eta$
(it is easy to see that if $\forall \delta: \abs{f(\delta) - g(\delta)} \leq K$,
then $\abs{\inf_{\delta} f - \inf_{\delta} g} \leq K$).

\begin{lem}\label{lem:approximation}
Given $\eta \geq 0$ and two sequences $u$ and $u'$ of utilities such that 
$\abs{u'_t - u_t} \leq (t+1) \cdot \eta$ for all $t \geq 0$, we have 
$\abs{\val(u',T) - \val(u,T)} \leq (T+1) \cdot \eta$. Analogously,
if $u'_t = u_t + (t+1) \cdot \eta$ for all $t \geq 0$, then
$\val(u',T) = \val(u,T) + (T+1) \cdot \eta$.
\end{lem}

We recall a result about Markov chains, which states that for Markov chains with only aperiodic recurrent
classes, the vector $\mu \cdot M^t$ converges to a steady-state vector $\pi$,
and the rate of convergence is bounded by an exponential in $n$~\cite[Theorem~4.3.7]{Gallager} 
(see Appendix~\ref{sec:rate-convergence} for detailed computation). For all $j \in V$:
$$ \abs{ (\mu \cdot M^t)_j - \pi_j } \leq K_1 \cdot K_2^t $$
where $K_1, K_2$ are constants with $K_2 < 1$, namely $K_2 = (1-\alpha^{n^2})^{1/{3n^2}}$ where $\alpha$
is the smallest non-zero probability in $M$ (i.e., $\alpha = \min \{ M_{ij}  \mid M_{ij}>0 \}$)
and $n$ is the number of vertices of $M$. 

For general Markov chains (with possibly periodic recurrent classes), we adapt the
above result as follows. Consider the set $\T$ of transient vertices, each recurrent class $C_1, C_2, \dots, C_l$
with their respective period $d_1, d_2, \dots, d_l$, and let $d = \lcm \{d_1, \dots, d_l \}$ 
be their least common multiple. Note that $d_i \leq n$ for all $1 \leq i \leq l$
and $d$ is at most the product of all prime numbers smaller than n, thus at most exponential in $n$~\cite{Erdos89}. 
Then $M^d$ can be viewed as the transition matrix
of a Markov chain with aperiodic recurrent classes, and thus $\mu \cdot M^{d\cdot t}$ 
converges to a steady-state vector $\pi$ as $t \to \infty$. 
Considering a recurrent class $C_i$, and the vertices $j \in C_i \cup \T$ 
the rate of convergence can be bounded as follows, where $\alpha^{d_i}$ is a lower bound on the smallest non-zero probability in $M^{d_i}$:
\begin{align*}
\abs{ (\mu \cdot M^{d \cdot t})_j - \pi_j } & = \abs{ (\mu \cdot (M^{d_i})^{\frac{d \cdot t}{d_i}})_j - \pi_j }  \\
& \leq K_1 \cdot (1-\alpha^{d_i \cdot n^2})^{\frac{d \cdot t}{d_i \cdot 3n^2}} \\
& \leq K_1 \cdot (1-\alpha^{n^3})^{\frac{t}{3n^2}}, 
\end{align*}
which is independent of $i$, and thus holds for all $j \in V$. Let $K_3 = (1-\alpha^{n^3})^{1/{3n^2}}$.

It follows that $\abs{ \mu \cdot M^{d \cdot t} \cdot w^\transpose - \pi \cdot w^\transpose } \leq n \cdot W \cdot K_1 \cdot K_3^t$ where $W = \norm{w}$ is the largest absolute weight in $w$. Then 
for all $\epsilon > 0$, for all $t \geq \frac{\ln(\frac{\epsilon}{n \cdot W \cdot K_1})}{\ln(K_3)} =: B$,
we have $\abs{ \mu \cdot M^{d \cdot t} \cdot w^\transpose - \pi \cdot w^\transpose } \leq \epsilon$,
and by the same reasoning with initial distributions $\mu \cdot M, \mu \cdot M^2,  \dots, \mu \cdot M^{d-1}$
we get $\abs{ \mu \cdot M^{d \cdot t + k} \cdot w^\transpose - \pi \cdot M^k \cdot w^\transpose } \leq \epsilon$
for all $0\leq k < d$.

Consider the sequence $u'$ defined by 
$$u'_t = \left\{\begin{array}{ll} u_t  & \text{ for all } t \leq d \cdot B \\
u_{d \cdot B} + \sum_{k=d \cdot B + 2}^{t} \pi \cdot M^{k \% d} \cdot w^\transpose  & \text{ for all } t > d \cdot B\end{array} \right.$$
where $k \% d$ is the remainder of the division of $k$ by $d$. 
Intuitively, $u'_t$ approximates $u_t$ after time $t = d \cdot B$ by considering 
the (expected) weight at time $t$ to be given by the limit (expected) weight at 
the steady-state distribution.

Then $\abs{u'_t - u_t} \leq (t+1) \cdot \epsilon$ for all $t \geq 0$,
and therefore  $\abs{\val(u',T) - \val(u,T)} \leq \epsilon \cdot (T+1)$ (by Lemma~\ref{lem:approximation}). 
The sequence $u'$ is an ultimately periodic sequence of the form $A.C^{\omega}$ 
where $\abs{A} = d \cdot B$ and $\abs{C} = d$. Hence the optimal value of $u'$ 
is given by Lemma~\ref{lem:val-cycle} and can be obtained by computing the 
first $d \cdot B+d$ terms of the sequence $u'$, the steady-state vector $\pi$, the number $d$, and
the average weight $M_C = \frac{S_C}{\abs{C}}$ where $S_C = \sum_{i=0}^{d-1} \pi \cdot M^i \cdot w^\transpose$.
This provides a way to compute an approximation with additive error $\epsilon$ of the optimal value of a Markov chain
in time $O(P(n) \cdot T \cdot B \cdot d)$ where $P(n)$ is a polynomial
in the size of the Markov chain (that accounts for matrix multiplication, steady-state
vector computation, etc.). 
Using the fact that $(1-\frac{1}{x})^x \in O(1)$,
and that $\ln\left(1-\frac{1}{x}\right) \in O(-1/x)$ (by Lemma~\ref{lem:math} in Appendix),
we obtain the bounds in Theorem~\ref{thm:approx-Skolem-hard}
in the special cases where $\alpha$ or $n$ is constant.

\begin{thm}\label{thm:approx-Skolem-hard}
The optimal expected value of a Markov chain with expected stopping time $T$
can be computed to an arbitrary level of precision~$\epsilon > 0$, in time 
$$O\left(P(n) \cdot T \cdot \frac{\ln\left(\frac{\epsilon}{n \cdot W }\right)}{\ln(K_3)} \cdot 2^{O(n)}\right)$$
where $K_3 = (1 - \alpha^{n^3})^{1/3n^2}$ and $P(\cdot)$ is a polynomial.

If $\alpha$ (the smallest non-zero probability) is constant, then the computation time is in 
$$O\left(\frac{P(n) \cdot 2^{O(n)}}{\alpha^{n^3}}\cdot T \cdot \ln \left(\frac{n \cdot W}{\epsilon}\right)\right) \text{ (as } n \to \infty).$$

If $n$ (the number of vertices) is constant, then the computation time is in 
$$O\left(\frac{1}{\alpha^{O(1)}}\cdot T \cdot \ln \left(\frac{W}{\epsilon}\right) \right) \text{ (as } \alpha \to 0).$$
\end{thm}

\begin{figure}[!tb]
  \begin{center}
    \hrule
      \input{figures/markov-chain-lower-bound.tex}
    \hrule
      \caption{A family of Markov chains for Proposition~\ref{pro:mc-lower-bound}.  \label{fig:mc-lower-bound}}
  \end{center}
\end{figure}
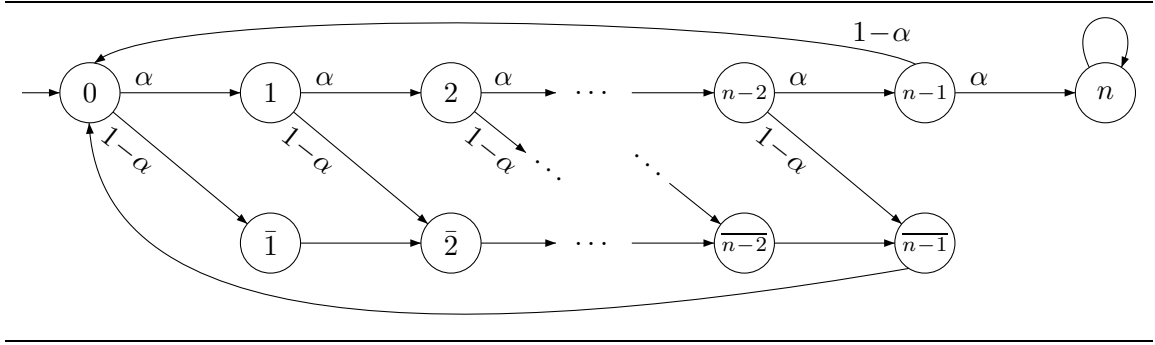

We present a lower bound on the execution time of the approximation algorithm of 
Theorem~\ref{thm:approx-Skolem-hard}: we show that the algorithm runs
in time exponential in the number of vertices of the Markov chains presented in 
\figurename~\ref{fig:mc-lower-bound}. This shows that the complexity analysis of 
our algorithm cannot be improved to eliminate the exponential dependency in the
number of vertices. However, whether there exists a polynomial-time algorithm for 
the approximation problem is an open question. 

\figurename~\ref{fig:mc-lower-bound} shows a family of Markov chains $M$ with $2n$
vertices and transition probabilities parameterized by $\alpha \leq \frac{1}{2}$,
with initial distribution $\mu$ that assigns probability~$1$ to vertex~$0$.
The steady-state distribution assigns probability~$1$ to vertex~$n$. It is easy
to show that from vertex~$0$ the probability to reach vertex~$n$ after $t\cdot n$ steps
is $1 - (1-\alpha^n)^t$. Therefore the distance between the distribution $\mu \cdot M^t$
at time $t$ and the steady-state vector is at least $(1-\alpha^n)^{\frac{t}{n}}$.
It follows that, to ensure that this distance is less than~$\epsilon$,
the algorithm needs to compute the sequence of utilities of the Markov chain up to time at least
$\frac{n\cdot \ln(\epsilon)}{\ln(1-\alpha^n)} \geq \frac{n \cdot \ln(1/\epsilon)}{\alpha^n}$
(using Lemma~\ref{lem:math} in Appendix to get $\frac{1}{\ln(1- \frac{1}{x})} \geq -x$ for all $x > 1$).

\begin{prop}\label{pro:mc-lower-bound}
There exists a family of aperiodic Markov chains $M(n,\alpha)$ with $2n$ vertices ($n\in \nat$) 
and smallest positive transition probability $\alpha$ ($\alpha \leq \frac{1}{2}$) such that,
for the initial distribution $\mu = (1,0,\dots,0)$, we have 
$$\max_{j} \, \abs{ (\mu \cdot M(n,\alpha)^t)_j - \pi_j } \geq (1-\alpha^n)^\frac{t}{n},$$
where $\pi$ is the steady-state vector of $M(n,\alpha)$,
and the computation time of the approximation algorithm (of Theorem~\ref{thm:approx-Skolem-hard}) for $M(n,\alpha)$
is at least $$\frac{n \cdot \ln(1/\epsilon)}{\alpha^n}.$$
\end{prop}

\section{Markov Decision Processes}

Markov decision processes (MDPs) extend Markov chains with transition choices 
determined by control actions. We give the basic definitions of MDPs and of
the optimal expected value of an MDP with expected stopping time $T$.

\subsection{Definitions}
A \emph{Markov decision process} is a tuple $\M = \tuple{V,A,\theta,\mu,w}$ consisting of:
\begin{itemize}
\item a finite set $V$ of vertices and a finite set $A$ of actions,
\item a transition function $\theta: V \times A \to (V \to [0,1])$
such that $\theta(v,a)$ is a probability distribution over~$V$,
that is $\sum_{v' \in V} \theta(v,a)(v') = 1$ for all $v \in V$ and $a \in A$.
\item $\mu: V \to [0,1]$ is an initial distribution 
and $w: V \to \rat$ is a vector of weights, as in Markov chains. 
\end{itemize}

Given a vertex $v \in V$ and a set $U \subseteq V$,
let $A_U(v)$ be the set of all actions $a \in A$ such 
that $\Supp(\theta(v,a)) \subseteq U$. 
A \emph{closed} set in an MDP is a set $U \subseteq V$ such 
that $A_U(v) \neq \emptyset$ for all $v \in U$. 
A set $U \subseteq V$ is an {\em end-component}~\cite{deAlfaro97,BK08} if 
(i)~$U$ is closed, and 
(ii)~the graph $(U,E_U)$ is strongly connected
where $E_U=\{(v,v') \in U \times U \mid \theta(v,a)(v') > 0 \text{ for some } a \in A_U(v)\}$
denote the set of edges given the actions. 
In the sequel, end-components should be considered \emph{maximal}, that is 
such that no strict superset is an end-component.

A \emph{strategy} in $\M$ is a function $\straa: V^+ \to (A \to [0,1])$ such that
$\straa(\rho)$ is a probability distribution over~$A$, for all sequences $\rho \in V^+$.
A strategy $\straa$ is \emph{pure} if for all $\rho \in V^+$, there exists an action $a \in A$
such that $\straa(\rho)(a) = 1$; $\straa$ is \emph{memoryless} if $\straa(\rho v) = \straa(\rho' v)$
for all $\rho,\rho' \in V^*$ and $v \in V$; $\straa$ uses \emph{finite memory}
 if there exists a right congruence $\approx$ over $V^+$ (i.e., 
 if $\rho \approx \rho'$, then $\rho \cdot v  \approx \rho' \cdot v$
 for all $\rho, \rho' \in V^+$ and $v \in V$)
 of finite index such that $\rho \approx \rho'$ implies $\straa(\rho) = \straa(\rho')$.

Given the initial distribution $\mu$, and a strategy $\straa$,
a probability can be assigned to every finite path $\rho = v_0  \cdots v_n$
as follows: 
$$\P_{\mu}^{\straa}(v_0 v_1 \dots v_k) = \mu(v_0) \cdot \prod_{i=0}^{k-1} \sum_{a \in A} \straa(v_0 \cdots v_{i})(a) \cdot \theta(v_{i}, a)(v_{i+1}).$$
Analogously, we denote by $\E_{\mu}^{\straa}(f)$ the expected value of 
the function $f: V^* \to \rat$ defined over finite sequences of vertices.
Let $u_t = \E_{\mu}^{\straa}( \sum_{i=0}^{t} w(v_i))$ and define
the optimal expected value of $\M$ with expected stopping time $T \in \rat$ as follows:
$$ \val(\M,T) = \sup_{\straa} \inf_{\substack{\delta \in \Delta \\ \E_{\delta} = T}} \sum_{t=0}^{\infty} \delta(t) \cdot u_t.$$
The strategy $\straa$ is \emph{$\epsilon$-optimal} if the sequence $u=(u_t)_{t \in \nat}$ it induces is such that 
$ \val(u,T) \geq \val(\M,T) - \epsilon$. For $\epsilon=0$, we simply say that $\straa$ 
is \emph{optimal} (instead of $0$-optimal).

For an arbitrary strategy $\straa$, with probability~$1$ the set of states visited 
infinitely often along an (infinite) path is an end-component~\cite{CY95,deAlfaro97}.
Let the limit-probability of a (maximal) end-component~$U$ be the probability that
the set of states visited infinitely often along a path is a subset of $U$.
A \emph{limit distribution} under $\straa$ is a distribution $\delta^*$ such that,
for every end-component~$U$,  the limit-probability of $U$ is $\sum_{v \in U} \delta^*(v)$.


\subsection{Infinite memory is necessary}

Since MDPs are an extension of Markov chains, the problem of computing the optimal expected value $\val(\M,T)$
is Positivity-hard (by Corollary~\ref{cor:pos-to-exact}). Another source of hardness
for this problem is that infinite memory is required for optimal strategies, 
as illustrated in the following example.

\begin{figure}[!t]
  \begin{center}
    \hrule
      \input{figures/mdp-infinite-mem.tex}
    \hrule
      \caption{An MDP where infinite memory is required for optimal expected value. \label{fig:mdp-infinite-mem}}
  \end{center}
\end{figure}
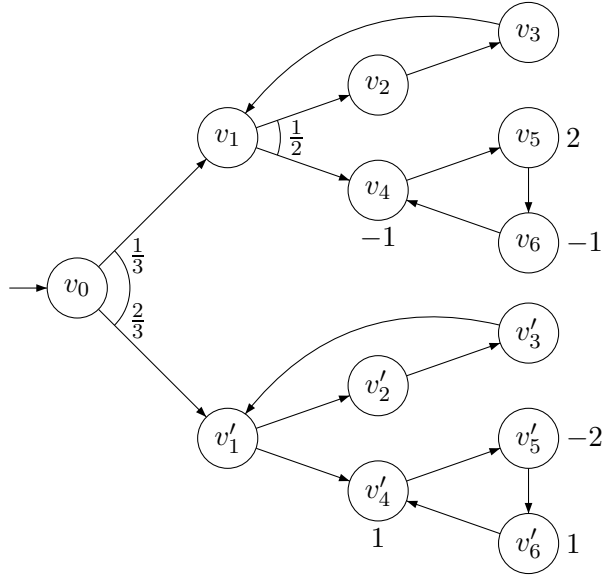

\medskip\noindent{\em Example.}
We show in \figurename~\ref{fig:mdp-infinite-mem} an MDP where infinite memory 
is required for optimal expected value.
The only strategic choice is in vertex $v'_1$ (we omit the actions in the figure, and all weights not shown are $0$). 
In particular, the upper part
$\{v_1,\dots,v_6\}$ is a Markov chain and after $3k+2$ steps,
the probability mass in $v_4$ is $p_k = \frac{1}{3}\cdot (1-\frac{1}{2^{k+1}})$.
For instance $p_0 = \frac{1}{6}$. Note that one step before, the probability mass in $v_1$ is $\frac{1}{3} \cdot \frac{1}{2^{k}}$.

We claim that the optimal expected value of the MDP is $0$,
which can be obtained by a strategy $\straa_{{\rm opt}}$ 
that ensures utility $0$ at every step:
let $m_k$ be the mass of probability in $v'_1$ after $3k+1$ steps (thus $m_0 =
\frac{2}{3}$, and $m_1, m_2, \dots$ depend on the strategy).
In $v'_1$, after $3k+1$ steps, the strategy $\straa_{{\rm opt}}$ 
chooses $v'_4$ with probability $\alpha_k$
such that $m_0 \cdot \alpha_0 = p_0$, thus $\alpha_0 = \frac{1}{4}$,
and $m_k \cdot \alpha_k = p_k - p_{k-1}$ for all $k \geq 1$.
It is easy to see that $m_k = \frac{1}{3} + \frac{1}{3} \cdot \frac{1}{2^{k}}$
and $\alpha_k = \frac{1}{2 + 2^{k+1}}$ ensure this 
as well as $m_{k+1} = m_{k} \cdot (1-\alpha_k)$ for all $k \geq 0$.
Therefore the strategy $\straa_{{\rm opt}}$ maintains always the
same probability in $v'_4$ as in $v_4$, and the expected total reward is
$0$ at every step.

It is easy to show that any other strategy (with a different value of some $\alpha_k$)
produces a negative total utility 
at some time step (either by putting too much probability into $v'_4$, and 
thus too much probability for weight $-2$ in $v'_5$, as compared to
the weight $2$ in $v_5$, or by putting too little probability into $v'_4$,
and thus too little probability for weight $1$ in $v'_4$, as compared to
the weight $-1$ in $v_4$), and that it entails a negative expected value of the MDP.

The strategy $\straa_{{\rm opt}}$ requires infinite memory, since 
the sequence $\alpha_k$ is strictly decreasing, and
the vertex $v'_1$ is reached after $3k+1$ steps along a unique path
$\rho_k = v_0v'_1(v'_2v'_3v'_1)^k$. It follows that for all
right congruences $\approx$ over $V^+$ such that $\rho \approx \rho'$ 
implies $\straa_{{\rm opt}}(\rho) = \straa_{{\rm opt}}(\rho')$, 
we have $\rho_k \not\approx \rho_l$ for $k\neq l$ since 
$\alpha_k\neq \alpha_l$ for $k\neq l$, thus $\approx$ cannot have finite index.
\medskip

As the above example illustrates, infinite-memory strategies are required in MDPs.
The expected stopping-time problem can be formulated as a game between a player
that controls the transition choice and the opponent that chooses the stopping times.
However, the game is not a perfect-information game as the opponent chooses the 
stopping times without knowing the execution of the MDP
(in particular, the stopping-time distribution cannot be adapted
according to the outcome of the probabilistic choices in the MDP).
As a consequence, while finite-memory strategies are sufficient in finite-horizon 
planning (even in perfect-information stochastic games), in contrast we show
infinite-memory strategies are required.
In general, in imperfect-information probabilistic models such as probabilistic automata~\cite{Paz-Book,RabinProb63,Reisz99},
infinite-memory strategies are required~\cite{BGB12}, and the basic computational problems 
(such as optimal reachability probability) as well as their \emph{approximation} are undecidable~\cite{MHC03}.
However, our setting only represents limited imperfect information for the opponent,
and we establish in the rest of this section that the approximation problem is decidable.

\subsection{Approximation of the optimal value}
The problem of computing $\val(\M,T)$ up to an additive error $\epsilon$ can be solved
as follows. We show that there exist $\epsilon$-optimal strategies of a simple form:
after some time $t^*$ (that depends on $\epsilon$), it is sufficient to play a (memoryless)
strategy that maximizes the mean-payoff expected reward, defined as follows
for a strategy $\straa$ in $\M$:
$$ \MP(\M,\straa) = \limsup_{t \to \infty} \frac{1}{t} \sum_{i=0}^{t-1} \E_{\mu}^{\straa}(w(v_i)),$$
and the \emph{optimal mean-payoff value} is
$$ \val^{\MP}(\M) = \sup_{\straa}  \MP(\M,\straa).$$

\begin{rem}\label{rmk:mdp-folklore}
It is known that (see e.g.~\cite{Puterman}):
\begin{itemize}
\item pure memoryless strategies are sufficient for mean-payoff optimality,
that is there exists a pure memoryless strategy $\straa$ such that $\val^{\MP}(\M) = \MP(\M,\straa)$; 
\item for variants of the definition of mean-payoff expected reward (using $\liminf$ instead of $\limsup$), 
or where the $\limsup$ and $\E(\cdot)$ operators are swapped (also known as the expected mean-payoff 
value), the same pure memoryless strategy is optimal;
\item all vertices in an end-component have the same optimal mean-payoff value.
\end{itemize}
\end{rem}

\begin{figure}[!tb]
  \begin{center}
    \hrule
      \input{figures/proof-sequence.tex}
    \hrule
      \caption{Main steps towards the proof that the supremum of total expected reward is bounded in MDPs with mean-payoff value at most~$0$ (Theorem~\ref{theo:sup-mdp}). \label{fig:proof-sequence}}
  \end{center}
\end{figure}
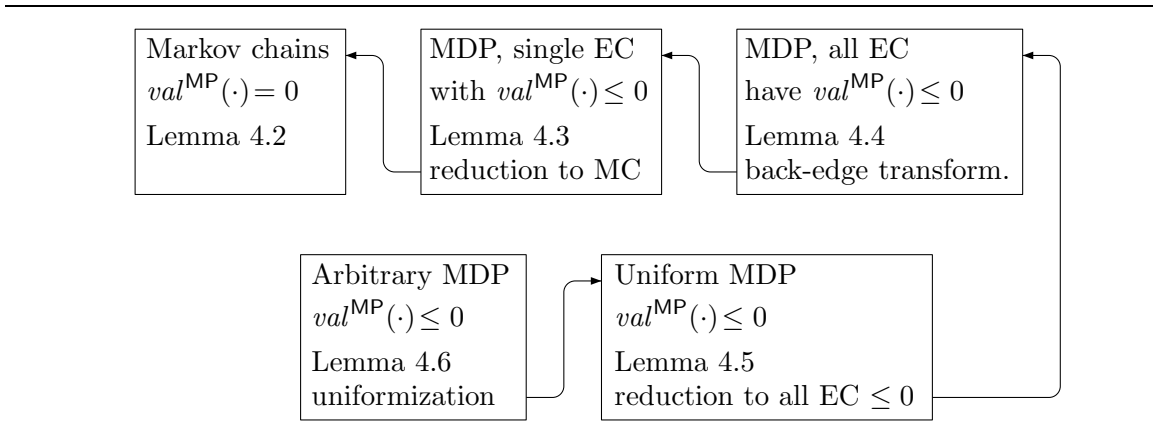

Intuitively, a strategy $\sigma$ that plays according to an optimal mean-payoff strategy after some time $t^*$
has an asymptotic behaviour that is at least as good as any strategy, in particular 
any $\epsilon$-optimal strategy; up to time $t^*$ (thus for finitely many steps), 
if the strategy $\sigma$ plays like an $\epsilon$-optimal strategy, then the sequence
of expected reward (defined above as $u_t$) is also good enough; the only question
is whether switching to an optimal mean-payoff strategy may induce a transient loss 
of reward after $t^*$ that could impede $\epsilon$-optimality. In fact, we show that 
$(1)$ the loss is bounded, and $(2)$ the impact of a bounded loss on the expected 
value is negligible if $t^*$ is large enough. 
That the loss is bounded, namely: 
$$ \sup_{\straa} \sup_{t} \, \sum_{i=0}^{t} \E_{\mu}^{\straa}(w(v_i)) \text{ is bounded if } \val^{\MP}(\M) \leq 0, $$ 
may appear intuitively true, but is not simple to prove
even in the special case where the mean-payoff value is $0$. 
The proof has several steps, summarized in \figurename~\ref{fig:proof-sequence},
leading to Theorem~\ref{theo:sup-mdp}.
We start by proving that the loss is bounded in the simple case of Markov chains
with mean-payoff value~$0$, then for larger classes of MDPs, using reductions
that transform an MDP $\M$ of a larger class into an MDP $\M'$ of a smaller class for which
a bound on the loss is already established. The transformations may increase the
total expected reward (as then, an upper bound for $\M'$ gives an upper bound for $\M$).

\begin{lem}\label{lem:sup-mc}
In aperiodic Markov chains $\tuple{M,\mu,w}$ with smallest positive transition 
probability $\alpha$, if the mean-payoff value, defined as
$\limsup_{t \to \infty} \frac{1}{t} \sum_{i=0}^{t-1} \E(w(v_i))$, is $0$,
then 
$$ \sup_{t} \, \left\lvert \sum_{i=0}^{t} \E(w(v_i)) \right\lvert \leq 4nW \cdot t_0 $$
where $t_0 = 3 \cdot n^5 \cdot \big(\frac{1}{\alpha}\big)^{n^2}$,
and $W$ is the largest absolute weight according to $w$.
\end{lem}

\begin{proof}
Consider the convergence rate of aperiodic Markov chains (Appendix~\ref{sec:rate-convergence}):
$$ \abs{ (\mu \cdot M^t)_j - \pi_j } \leq 3 \cdot K^t \quad \text{ for all } j \in V $$
where $K$ is a constant with $K < 1$, namely $K = (1-\alpha^{n^2})^{1/{3n^2}}$. 

Consider time $t_0 = 3 \cdot n^5 \cdot \big(\frac{1}{\alpha}\big)^{n^2}$, which
is such that $K^{t_0} \leq \frac{1}{2^{n^3}}$ for all $n \geq 1$ (using Lemma~\ref{lem:math} in Appendix).

Now consider the expected total reward at time $t$, given by 
$$ \left\lvert \sum_{i=0}^{t} \E(w(v_i)) \right\lvert =  \left\lvert \sum_{i=0}^{t} \mu \cdot M^i \cdot w \right\lvert $$

and show that it is bounded by $4nW \cdot t_0$, for all $t \geq 0$.
The proof goes by showing a bound on the total reward that can be accumulated within a time unit.
At time $i$, the reward per time unit is $\mu \cdot M^i \cdot w \leq 3nW \cdot K^i$,
since the mean-payoff value of the Markov chain is~$0$, which implies $\pi \cdot w = 0$.
For times $t < t_0$, we bound the total reward per time unit trivially by $W$.
For times $k\cdot t_0 \leq t < (k+1) \cdot t_0$ (where $k\geq 1$), we bound the total 
reward per time unit by $3nW \cdot \big(\frac{1}{2^{n^3}}\big)^{k}$ since 
$3\cdot K^t \leq 3\cdot K^{k\cdot t_0} \leq 3 \cdot \big(\frac{1}{2^{n^3}}\big)^{k}$.

It is now sufficient to show that the sum of the bounds on the total reward per time unit
is bounded by $t_0 \cdot (W+1)$ for arbitrarily large $t$, which we establish as follows:

\begin{equation}
t_0 \cdot W + \sum_{k=1}^{\infty} 3nW \cdot \big(\frac{1}{2^{n^3}}\big)^{k} \cdot t_0 = t_0 \cdot W + 3nW\cdot \frac{t_0}{2^{n^3}-1} \leq 4nW \cdot t_0. \tag*{\qed}
\end{equation}
\renewcommand{\qedsymbol}{}
\end{proof}

To prove a similar result for MDPs (Theorem~\ref{theo:sup-mdp}), we first consider the 
case of MDPs that consist of a single end-component, and show by contradiction that
if it has mean-payoff value $0$ and a large expected total reward could be accumulated 
from a vertex $v_0$ using some strategy $\sigma_0$, 
then by reaching $v_0$ again (which is possible since the MDP is strongly connected) and 
repeating the same strategy $\sigma_0$, we could get a strictly positive mean-payoff value.
A technical difficulty in this proof is that $v_0$ may be reached by paths of different
lengths, but the large expected total reward that can be accumulated from $v_0$ is obtained
in a fixed number of steps. 

\begin{lem}\label{lem:sup-mdp-ec}
In an MDP~$\M$ that is an end-component (i.e., $V$ is an end-component), if $\val^{\MP}(\M) \leq 0$ and $\abs{V} = n$, then 
$$ \sup_{\straa} \sup_{t} \, \sum_{i=0}^{t} \E_{\mu}^{\straa}(w(v_i)) \leq 12\cdot n^6 \cdot W \cdot \bigg(\frac{1}{\alpha}\bigg)^{n^3}$$
where $\alpha$ is the smallest positive transition probability in $\M$,
and $W$ its largest absolute weight.
\end{lem}

\begin{proof}
Assume towards contradiction that there exists an initial distribution $\mu$
such that the inequality of the lemma does not hold in $\M$. 
It follows that in some initial vertex $v_0$ (such that $\mu(v_0) > 0$) the inequality does not hold, 
i.e. there exists a strategy $\straa_0$ and time $t_0$ such that the expected total reward from $v_0$ under $\straa_0$
at time $t_0$ is at least $12\cdot n^6 \cdot W \cdot (\frac{1}{\alpha})^{n^3}$. 

First we modify the MDP $\M$ to obtain an MDP $\M'$ as follows, 
in a way that does not decrease the value of $\sup_{\straa} \sup_{t} \, \sum_{i=0}^{t} \E_{\mu}^{\straa}(w(v_i))$~:
\begin{itemize}
\item increase the weight of every vertex by $\abs{\val^{\MP}(\M)}$ (define $w'(v) = w(v) - \val^{\MP}(\M)$), and
\item add a copy $\hat{v}_0$ of $v_0$ with weight $-W$ and a self-loop; formally,
define a new action set $A' = A \cup \{\hat{a} \mid a \in A\}$, a new state space $V' = V \cup \{\hat{v_0}\}$
with $w'(\hat{v_0}) = -W$ and $w'(v) = w(v)$ for all $v \in V$, and transitions on $\hat{a}$ 
that replace $v_0$ by $\hat{v}_0$ as follows, for all $v \in V$ and all $a \in A$:
$\theta'(v,\hat{a})(\hat{v}_0) = \theta(v,a)(v_0)$, $\theta'(v,\hat{a})(v_0) = 0$, and 
$\theta'(v,\hat{a})(v') = \theta(v,a)(v')$ for all $v' \in V\setminus \{v_0\}$.
The self-loop on $\hat{v}_0$ is defined on all actions $\hat{a}$ (i.e., $\theta'(\hat{v}_0,\hat{a})(\hat{v}_0)=1$), 
and the other actions have the same effect as from $v_0$ (i.e., $\theta'(\hat{v}_0,a) = \theta(v_0,a)$
and $\theta'(v,a) = \theta(v,a)$ for all $v\in V$).
\end{itemize}

In the new MDP $\M' = \tuple{V',A',\theta',\mu,w'}$, we note that:
\begin{itemize}
\item the expected total reward from $v_0$ is not smaller in $\M'$ than in $\M$,
since we increased weights of existing transitions, and we added new transitions,
which cannot decrease the expected total reward (strategies of $\M$ can still be
played in $\M'$);
\item the optimal mean-payoff value of $\M'$ is $\val^{\MP}(\M') = 0$ since
increasing all weights by $\abs{\val^{\MP}(\M)}$ has the effect to increase
the mean-payoff value by the same amount; moreover, adding the copy $\hat{v}_0$
with weight $-W$ does not change the optimal mean-payoff value. To see this, 
fix a memoryless strategy $\straa$, and consider the recurrent classes of the resulting
Markov chain. If a recurrent class $C$ contains $\hat{v}_0$, then either
the self-loop on $\hat{v}_0$ is used by the strategy $\straa$ and then the mean-payoff 
value of $C$ is $-W$, or the self-loop on $\hat{v}_0$ is not used and the mean-payoff 
value of $C$ is less than the mean-payoff value of $C' = C \cup {v_0} \setminus \{\hat{v}_0\}$
which is a recurrent class that can be obtained in $\M$ using the strategy that copies
$\straa$ but plays $a$ whenever $\straa$ plays $\hat{a}$. Since $\val^{\MP}(\M) \leq 0$,
it follows that the mean-payoff value of $C'$ (and thus of $C$) is at most $0$.
On the other hand, if $C$ does not contain $\hat{v}_0$, then it can be
obtained in $\M$ and thus its mean-payoff value is also at most $0$.
\item the state space of $\M'$ is of size $\abs{V'} = n+1 \leq n^2$,
and $\M'$ is still an end-component.
\end{itemize}

Given the strategy $\straa_0$ and time $t_0$ as above, we show that there exists
a strategy $\straa^*$ and a time $t^*$ such that from all vertices $v \in V$, we have 
\begin{equation}
  \sum_{i=0}^{t^*-1} \E_{v}^{\straa^*}(w'(v_i)) \geq 5 \cdot n^6 \cdot W \cdot \bigg(\frac{1}{\alpha}\bigg)^{n^3} > 0 \label{eq:bound}
\end{equation}
which entails, since the bound $t^*$ is the same for all vertices, that 
by repeating the strategy $\straa_0$ every $t_0$ steps
the mean-payoff value of $\M'$ is positive, $\val^{\MP}(\M') > 0$, in contradiction
with the fact that $\val^{\MP}(\M') = 0$.

Let $t^* = t_{\reach} + t_0 $ where $t_{\reach} = \frac{n}{\alpha^{n}}$ 
and we construct $\straa^*$ that plays as follows:
\begin{itemize}
\item 
for $t_{\reach}$ steps, play a pure memoryless strategy $\straa_{\reach}$ to reach $\hat{v}_0$ almost-surely
(and play the self-loop with weight $-W$ in $\hat{v}_0$); such a strategy exists
because the MDP is an end-component~\cite{deAlfaro97};
\item
after $t_{\reach}$ steps: if the current vertex is $\hat{v}_0$, play for the next $t_0$ steps
the strategy $\straa_0$; if the  current vertex is not $\hat{v}_0$, play for the next $t_0$ steps
a memoryless optimal strategy for the mean-payoff value (thus using only actions in $A$ and
without visiting $\hat{v}_0$). 
\end{itemize}

The value $t_{\reach}$ is such that the probability mass in $\hat{v}_0$
after $t_{\reach}$ steps is at least $\frac{1}{2}$:
since the strategy $\straa_{\reach}$ is pure memoryless and $\abs{V'} = n+1$, we can use the
analysis of Markov chain reachability to claim that the probability to 
have reached target vertex~$\hat{v}_0$ after $t_{\reach} = k\cdot n$ steps is at least 
$1 - (1-\alpha^{n})^k > \frac{1}{2}$ since $k = \frac{1}{\alpha^{n}}$ and
$\big(1-\frac{1}{x}\big)^x < e^{-1} < \frac{1}{2}$ (by Lemma~\ref{lem:math} in Appendix).

We bound the expected total reward of $\straa^*$ as follows:
\begin{itemize}
\item after $t_{\reach}$ steps, since all weights are bounded by $W$, the expected total reward
is at least $- t_{\reach} \cdot W$;
\item in the next $t_0$ steps, the collected reward from $\hat{v}_0$ (in which the probability mass is at least $\frac{1}{2}$)
is at least $12\cdot n^6 \cdot W \cdot (\frac{1}{\alpha})^{n^3}$ (by the definition
of $\straa_0$ and $t_0$), and the collected reward from other vertices is at least
$-12\cdot n^6 \cdot W \cdot (\frac{1}{\alpha})^{n^2}$ (by Lemma~\ref{lem:sup-mc}, since the optimal strategy 
for the mean-payoff value is memoryless and plays only actions in $A$, which gives a Markov chain 
with $n$ vertices and mean-payoff value equal to~$0$).
\end{itemize}
It follows that the expected total reward of $\straa^*$ after $t^*$ steps is at 
least:
\begin{align*}
& - \frac{nW}{\alpha^n} + 6 \cdot n^6 \cdot W \cdot \bigg(\frac{1}{\alpha}\bigg)^{n^3} -  6\cdot n^6 \cdot W \cdot \bigg(\frac{1}{\alpha}\bigg)^{n^2} \\
& \geq 6 \cdot n^6 \cdot W \cdot \bigg(\frac{1}{\alpha}\bigg)^{n^3} -  7\cdot n^6 \cdot W \cdot \bigg(\frac{1}{\alpha}\bigg)^{n^2} \\
& \geq 5 \cdot n^6 \cdot W \cdot \frac{1}{\alpha^{n^3}} \ \text{ since } n \geq 2 \text{ and } \alpha \leq \frac{1}{2}, 
\end{align*}
which establishes the bound~\eqref{eq:bound} and concludes the proof.
\end{proof}

We can easily extend the result to MDPs with several end-components, if all of them 
have mean-payoff value at most $0$.

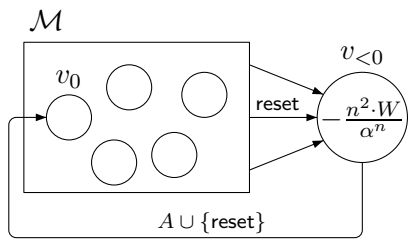
\begin{figure}[!tb]
  \begin{center}
    \hrule
      \input{figures/mdp-reset.tex}
    \hrule
      \caption{Back-edge transformation (Proof of Lemma~\ref{lem:sup-mdp-several-ec}). \label{fig:mdp-reset}}
  \end{center}
\end{figure}

\begin{lem}\label{lem:sup-mdp-several-ec}
In an MDP~$\M$ with $n$ vertices in which all end-components have an optimal mean-payoff value at most~$0$, we have 
$$ \sup_{\straa} \sup_{t} \, \sum_{i=0}^{t} \E_{\mu}^{\straa}(w(v_i)) \leq 12\cdot n^8 \cdot W \cdot \bigg(\frac{1}{\alpha}\bigg)^{n^3+n}$$
where $\alpha$ is the smallest positive transition probability in $\M$,
and $W$ its largest absolute weight.
\end{lem}

\begin{proof}
Consider an MDP~$\M$ as in the lemma statement, and assume without loss of generality
that the initial distribution~$\mu$ is a Dirac distribution, namely $\mu(v_0) = 1$ 
for some vertex $v_0$.

We present the back-edge transformation from $\M$ to an MDP $\M'$ as follows (\figurename~\ref{fig:mdp-reset}).
We add a special action $\reset$, thus the action set of $\M'$ is $A \cup \{\reset\}$. 
The state space of $\M'$ is $V \cup \{v_{<0}\}$
where $v_{<0}$ is a new vertex with weight $-\frac{n^2 \cdot W}{\alpha^n}$, 
and the transition functions of $\M'$ and $\M$ are identical for actions in $A$ and vertices in $V$.
On action $\reset$, the transition function $\theta'$ of $\M'$ has a ``back-edge'' 
from every vertex to $v_{<0}$; from $v_{<0}$ there is an edge to $v_0$ on every action.
Thus $\theta'(v,\reset) = v_{<0}$ for all vertices $v \in V$, and
$\theta'(v_{<0},a) = \mu$ for all actions $a \in A \cup \{\reset\}$.

First note that the supremum of expected total reward is not smaller in $\M'$
than in $\M$, since $\M'$ contains all vertices and transitions of $\M$.

We now show that the mean-payoff value of $\M'$ is at most $0$,
which establishes the bound in the lemma statement as follows:
since the whole $\M'$ is an end-component, we can apply Lemma~\ref{lem:sup-mdp-ec}
where the largest absolute weight in $\M'$ is $\frac{n^2 \cdot W}{\alpha^n}$, which gives 
the announced bound for $\M'$, thus also for $\M$.

To show that the mean-payoff value of $\M'$ is at most $0$, fix an optimal
strategy $\straa$ for mean-payoff in $\M'$, which we can assume to be pure 
memoryless (Remark~\ref{rmk:mdp-folklore}). In the resulting Markov chain,
we show that all recurrent classes have mean-payoff value at most~$0$
as follows: if a recurrent class does not contain $v_{<0}$, then it is
an end-component in the original MDP $\M$, and therefore its mean-payoff value
is at most $0$; otherwise it contains $v_{<0}$ and since the frequency\footnote{The frequency of the
vertex with largest frequency is at least $f_n = \frac{1}{n+1}$, and the frequency 
of the $(k+1)$-th frequent vertex is at least $\alpha^k \cdot f_n$.}
$f_0$ of a recurrent vertex is at least $\frac{\alpha^n}{n+1} \geq \frac{\alpha^n}{n^2}$, the mean-payoff value
of the recurrent class is at most $- f_0 \cdot \frac{n^2 \cdot W}{\alpha^n} + (1-f_0) \cdot W \leq -W + W = 0$.
This shows that  all recurrent classes have mean-payoff value at most~$0$,
and thus the optimal mean-payoff value of the MDP $\M'$ is at most $0$.
\end{proof}

\begin{figure}[!tb]
  \begin{center}
    \hrule
      \input{figures/mdp-pos-neg.tex}
    \hrule
      \caption{An MDP with positive and negative end-components. Its mean-payoff value is $0$. \label{fig:mdp-pos-neg}}
  \end{center}
\end{figure}
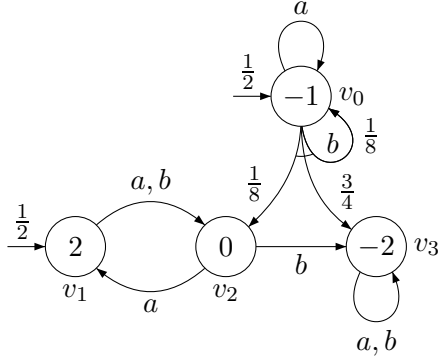

In an arbitrary MDP with mean-payoff value at most~$0$, some end-components 
may have positive value, and others negative value, as in the example 
of \figurename~\ref{fig:mdp-pos-neg}:
the three end-components $\{v_0\}$, $\{v_1,v_2\}$, $\{v_3\}$ have respective 
mean-payoff value $-1$, $1$, and $-2$. From the initial distribution $\mu$
where $\mu(v_0) = \mu(v_1) = \frac{1}{2}$, the mean-payoff value is $0$.
The case where the MDP has some end-components
with positive mean-payoff value requires a slightly more technical proof (see 
also \figurename~\ref{fig:proof-sequence}):
we first show in Lemma~\ref{lem:sup-mdp-ec-uniform} 
that the supremum of expected total reward in MDPs is bounded 
if all end-components are \emph{uniform} (an end-component is uniform if all 
its vertices have the same weight); then we present uniformization in Lemma~\ref{lem:sup-mdp-transform} to transform arbitrary MDPs into uniform MDPs.  

Given an MDP with weight vector $w$, let $\calE$ be the union of all its
end-components. Define the vector $w_{\trans}$ and $w_{\ec}$ as follows:
\begin{align*}
 w_{\trans}(v) & = \left\{ \!\begin{array}{ll}w(v) & \text{if } v \in V \setminus \calE \\
                                           0    & \text{if } v \in \calE \\
   \end{array}  \right. \\
 w_{\ec}(v) & = \left\{ \!\begin{array}{ll}0    & \text{if } v \in V \setminus \calE \\
                                       w(v) & \text{if } v \in \calE \\
   \end{array}  \right.
\end{align*}

It follows that $w =  w_{\trans} + w_{\ec}$ and by the triangular inequality,
we have 
\begin{align*}
\sup_k \sum_{i=0}^{k} \E_{\mu}^{\straa}(w(v_i)) \leq \sup_k \sum_{i=0}^{k} \E_{\mu}^{\straa}(w_{\trans}(v_i)) + \sup_k \sum_{i=0}^{k} \E_{\mu}^{\straa}(w_{\ec}(v_i)).
\end{align*}
Using Lemma~\ref{lem:sup-mdp-several-ec}, it is easy to bound the supremum of expected total reward for $w_{\trans}$,
and we present a bound on the supremum of expected total reward for $w_{\ec}$ in uniform MDPs as follows.

\begin{lem}\label{lem:sup-mdp-ec-uniform}
Given an MDP $\M$ with $n$ vertices, let $w_{\trans}$ and $w_{\ec}$ be the weight vectors of the transient vertices and of the end-components,
respectively. We have
$$\sup_{\straa} \sup_t \sum_{i=0}^{t} \E_{\mu}^{\straa}(w_{\trans}(v_i)) \leq 12\cdot n^8 \cdot W \cdot \left(\frac{1}{\alpha}\right)^{n^3+n},$$
and if $\val^{\MP}(\M) \leq 0$ and all end-components of $M$ are uniform, then
$$ \sup_{\straa} \sup_{t} \, \sum_{i=0}^{t} \E_{\mu}^{\straa}(w_{\ec}(v_i)) \leq 12\cdot n^8 \cdot W \cdot \left(\frac{1}{\alpha}\right)^{n^3+n},$$
where $\alpha$ is the smallest positive transition probability in $\M$,
and $W$ its largest absolute weight.
\end{lem}

\begin{proof}
For the first part, 
the bound for $w_{\trans}$ is given by Lemma~\ref{lem:sup-mdp-several-ec}, since the MDP with weight
vector $w_{\trans}$ has all its end-component with mean-payoff value $0$.

For the second part, assuming $\val^{\MP}(\M) \leq 0$ and all end-components of $M$ are uniform, the bound for $w_{\ec}$ is established as follows.
First consider the MDP $\M$ with weight function $w'$ defined by $w'(v) = w(v)$
if $v \in \calE$, and  $w'(v) = -W$ otherwise (i.e., for transient vertices).
Note that $w' = w_{\ec} - w_{0}$ where $w_{0}(v) = 0$ if $v \in \calE$, and  $w_{0}(v) = W$ otherwise. 
In $\M$ with $w'$, we will show that: 
\begin{equation}
  \sup_{\straa} \sup_{t} \, \sum_{i=0}^{t} \E_{\mu}^{\straa}(w'(v_i)) \leq 0. 
  \label{eq:bound-to-show}
\end{equation}
To obtain the bound for $w_{\ec} = w' + w_{0}$ and conclude the proof, 
we use the triangular inequality which entails that 
$\sup_{\straa} \sup_{t} \, \sum_{i=0}^{t} \E_{\mu}^{\straa}(w_{\ec}(v_i)) \leq 0 + B$
where $B$ is the bound given by the first part of the lemma for $w_{0}$.

To show~\eqref{eq:bound-to-show}, we consider an arbitrary strategy $\straa$ and we show that 
for all $n \geq 0$, the expected reward at step $n$ is at most $0$, that is:
$$ \sum_{v \in V} \delta^{\straa}_n(v) \cdot w'(v) \leq 0$$
where $\delta^{\straa}_n$ is the vertex distribution of $\M$ after $n$ steps 
under strategy $\straa$. 

Given $\delta^{\straa}_n$ (as an initial distribution), consider a memoryless 
strategy $\straa^{\calE}$ such that end-components
are never left, defined as follows for all $v \in V$:
if $v \in \calE$, then $\straa^{\calE}(v)$ is an action to stay in the end-component of $v$ 
in the next step;
otherwise, $\straa^{\calE}(v)$ is an arbitrary action. Such a strategy $\straa^{\calE}$ exists 
by definition of end-components.

By the assumption that $\val^{\MP}(\M) \leq 0$ (with weight vector $w$ thus also with $w'$), 
it follows that the limit distributions 
$\delta^*$ satisfy $\sum_{v \in \calE} \delta^*(v) \cdot \eta(v) \leq 0$.

Then, within the distribution $\delta^{\straa}_n$, the probability mass $p_{\calE}$ in $\calE$
never leaves an end-component, and the probability mass $1-p_{\calE}$ in $V \setminus \calE$
eventually (in the limit) gets injected in $\calE$ (and then never leaves). 
The (future) contribution of the probability mass $1-p_{\calE}$
to the expected reward of limit distributions is bounded below by $-W$ 
(since $\eta(v) \geq -W$ for all $v \in \calE$,
where $\eta(v)$ is the mean-payoff value of $v$ and of the end-component 
containing $v$ since $\M$ is uniform).
It follows that: 
\begin{align*}
0 & \geq \sum_{v \in \calE} \delta^*(v) \cdot \eta(v) \\
       & \geq \sum_{v \in \calE} \delta^{\straa}_n(v) \cdot \eta(v) + \sum_{v \in V \setminus \calE} \delta^{\straa}_n(v) \cdot (-W) \\
       & = \sum_{v \in V} \delta^{\straa}_n(v) \cdot w'(v)
\end{align*}
which concludes the proof by entailing~\eqref{eq:bound-to-show}.
\end{proof}

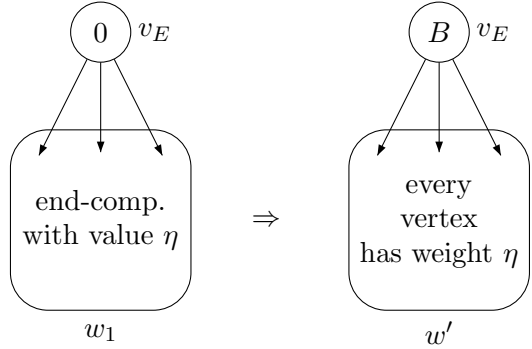
\begin{figure}[!tb]
  \begin{center}
    \hrule
      \input{figures/mdp-ec-bound.tex}
    \hrule
      \caption{Uniformization (including transformation of the weight vector $w_1$ into $w'$) for Lemma~\ref{lem:sup-mdp-transform}. The number $B$ is twice the bound given by Lemma~\ref{lem:sup-mdp-several-ec}. \label{fig:mdp-ec-bound}}
  \end{center}
\end{figure}

\medskip\noindent{\emph{Uniformization}.}
We present a \emph{uniformization} procedure that, given an MDP $\M$
with mean-payoff value at most~$0$, constructs an MDP $\M'$ with 
the same mean-payoff value as $\M$, with a larger supremum of expected total reward,
and in which all end-components are uniform. 

The procedure has two steps. First we construct $\M'$ and weight vector $w_1$ 
by transforming the structure of $\M$
in such a way that for each end-component, there is a single vertex from which
the end-component can be entered (as illustrated on the left of \figurename~\ref{fig:mdp-ec-bound}). 
This shape of MDP can be obtained as follows: for each end-component $E$, create
a new vertex $v_E$ (with weight~$w_1(v_E)=0$) with an edge from $v_E$ to every vertex in $E$, and modify
the transition function from every vertex $v$ outside $E$ to redirect all the 
probability mass that was going from $v$ to $E$ to go to $v_E$. Analogously
we transfer the probability of the initial distribution that was in $E$ to $v_E$.
By doubling the weight of every vertex to define $w_1$ (and inserting, for every vertex, 
a new intermediate vertex with weight~$0$ that is entered before going to the original vertex), 
it is easy but tedious to show that the mean-payoff value according to $w_1$ remains
the same, and that the supremum of expected total reward is at least twice the 
original one (note also that the number of vertices has at most tripled).

In the second step, we construct a new vector $w'$ of weights for $\M'$ 
such that every end-component becomes uniform (illustrated 
on the right of \figurename~\ref{fig:mdp-ec-bound}).
Given an end-component $E$ and the vertex $v_E$, let $w'(v_E) = B$ and $w'(v) = \eta$ where $\eta$
is the mean-payoff value of $E$ (according to $w_1$, or equivalently according to $w$)
and $B = 12\cdot n^8 \cdot (3W) \cdot \left(\frac{1}{\alpha}\right)^{n^3+n}$ is 
three times the bound given by Lemma~\ref{lem:sup-mdp-several-ec}.

\begin{lem}\label{lem:sup-mdp-transform}
Given an MDP~$\M$ with $n$ vertices, we can construct an MDP $\M'$ with the following properties:
\begin{enumerate}
\item all end-components of $\M'$ are uniform.\label{item1}
\item $\M$ and $\M'$ have the same mean-payoff value;\label{item2}
\item the number of vertices in $\M'$ is at most $3n$;\label{item3}
\item the supremum of expected total reward in $\M$ is less than half 
the supremum of expected total reward in $\M'$;\label{item4}
\end{enumerate}
\end{lem}

\begin{figure}[!tb]
  \begin{center}
    \hrule
      \input{figures/mdp-ec-bound-difference.tex}
    \hrule
      \caption{Over-approximation of the MDP with weight vector $w_{\dif} = w-w'$ (from \figurename~\ref{fig:mdp-ec-bound}). 
\label{fig:mdp-ec-bound-difference}}
  \end{center}
\end{figure}
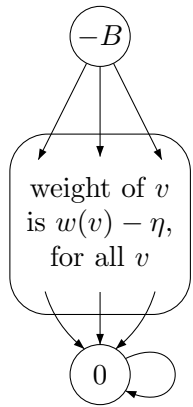

\begin{proof}
Consider $\M'$ obtained from $\M$ by the uniformization procedure.
Item~\ref{item1}.~holds by construction, and the proof of item~\ref{item2}.~and item~\ref{item3}.~has 
been sketched along with the uniformization procedure (note that the definition of $w'$ does not
change the mean-payoff value of the end-components, as compared to weight vector $w_1$ and $w$).

We show item~\ref{item4}.~for the transformation of one end-component~$E$ 
(\figurename~\ref{fig:mdp-ec-bound}), 
and the lemma follows by applying the result successively to each end-component.

Given an arbitrary strategy $\straa$, let $s_k(v) = \sum_{i=0}^{k} \E_{v}^{\straa}(w_1(v_i))$ and 
$s'_k(v) = \sum_{i=0}^{k} \E_{v}^{\straa}(w'(v_i))$
be the expected total reward from initial vertex $v$ after $k$ steps, 
according to the weight vector $w_1$ and $w'$ respectively.
We show that for all $v \in V \setminus E$, and for all $k$, we have 
$s_k(v) \leq s'_k(v)$, which establishes the claim that $\M'$ (with $w'$) has a larger 
supremum of expected total reward than $\M'$ (with $w_1$), since the initial distribution 
of $\M'$ has support in  $V \setminus E$ (and we showed that the supremum of expected total 
reward in $\M'$ (with $w_1$) is at least twice the one in $\M$). 
We show this by induction on $k$.
The base case $k=0$ holds since $w_1(v) \leq w'(v)$ for all $v \in V \setminus E$.
For the induction case, assume $s_t(v) \leq s'_t(v)$ for all $v \in V \setminus E$, and for all $t \leq k-1$.
Let $v \in V \setminus E$ and we consider two cases:

\begin{itemize}
\item
If $v \neq v_E$, the claim $s_k(v) \leq s'_k(v)$ holds since all successors
of $v$ are in $V \setminus E$ and we can use the induction hypothesis:
\begin{align*}
s_k(v) & = w_1(v) + \max_{a \in A} \sum_{u \in V \setminus E} s_{k-1}(u) \cdot \theta(v,a)(u) \\
       & \leq w_1(v) + \max_{a \in A} \sum_{u \in V \setminus E} s'_{k-1}(u) \cdot \theta(v,a)(u) \\
       & = s'_k(v)
\end{align*}

\item If $v = v_E$, then all successors of $v$ belong to $E$, and we cannot directly 
use the induction hypothesis. Consider the weight vector $w_{\dif} = w_1 - w'$,
and show that for $s^{\dif}_k(v) = \sum_{i=0}^{k} \E_{v}^{\straa}(w_{\dif}(v_i))$,
we have $s^{\dif}_k(v) \leq 0$, which implies $s_k(v) \leq s'_k(v)$.
Given the weight vector $w_{\dif}$, we know that starting from $v = v_E$, as
soon as a path leaves the end-component~$E$, its contribution to the expected
total reward is at most~$0$ (by induction hypothesis, since at most $k-1$
steps remain after exiting). Therefore, it is sufficient to show that
the expected total reward in $k$ steps is at most $0$ in the MDP of 
\figurename~\ref{fig:mdp-ec-bound-difference} where the edges going out of the
end-component $E$ are directed to a sink with weight $0$.
The weights defined by $w_{\dif}$ in $E$ are bounded by $\norm{w_1} +  \norm{w'} \leq 2W + W$.
The number of vertices in $E$ is less than $2n$ (where $n$ is the number of vertices in 
the original MDP $\M$), and only half of them are relevant to define the expected total reward.
By Lemma~\ref{lem:sup-mdp-several-ec}, since all end-components have mean-payoff value $0$ 
in the MDP of \figurename~\ref{fig:mdp-ec-bound-difference}, and~$B$
is the bound given by Lemma~\ref{lem:sup-mdp-several-ec}, we have 
$\sup_k s^{\dif}_k(v) \leq 0$ and therefore $s^{\dif}_k(v) \leq 0$,
which concludes the proof of the induction case. \qedhere
\end{itemize}
\end{proof}

We finally obtain an analogue of Lemma~\ref{lem:sup-mc} for MDPs: 
the expected total reward is bounded in MDPs with non-positive 
mean-payoff value. 

\begin{thm}\label{theo:sup-mdp}
Given an MDP~$\M$ with $n$ vertices and $\val^{\MP}(\M) \leq 0$, we have:
$$ \sup_{\straa} \sup_{t} \, \sum_{i=0}^{t} \E_{\mu}^{\straa}(w(v_i)) \in O\left( n^{16} \cdot W \cdot \left(\frac{1}{\alpha}\right)^{O(n^3)} \right) $$
where $\alpha$ is the smallest positive transition probability in $\M$,
and $W$ its largest absolute weight.
\end{thm}

\begin{proof}
Given MDP $\M$, we use the triangular inequality to bound the supremum
of expected total reward by the sum the supremum on the transient vertices
and on the end-components. For transient vertices, we use directly the bound 
in Lemma~\ref{lem:sup-mdp-ec-uniform},
and for end-components, we use the construction of $\M'$ in Lemma~\ref{lem:sup-mdp-transform},
and then apply Lemma~\ref{lem:sup-mdp-ec-uniform} where the number of vertices is $3n$,
and the largest weight is $B = 12\cdot n^8 \cdot (3W) \cdot \left(\frac{1}{\alpha}\right)^{n^3+n}$.
Since the supremum of expected total reward in $\M'$ is twice the 
supremum of expected total reward in $\M$, we get the following bound:
\begin{align*}
& \sup_{\straa} \sup_{t} \, \sum_{i=0}^{t} \E_{\mu}^{\straa}(w(v_i))  \\
& \leq 12\cdot n^8 \cdot W \cdot \left(\frac{1}{\alpha}\right)^{n^3+n} \\
& \phantom{\leq\, } + \frac{12}{2} \cdot (3n)^8 \cdot \left(12\cdot n^8 \cdot (3W) \cdot \left(\frac{1}{\alpha}\right)^{n^3+n}  \right) \cdot \left(\frac{1}{\alpha}\right)^{27n^3+3n} \\
& = \underbrace{12\cdot n^8 \cdot W \cdot \left(\frac{1}{\alpha}\right)^{n^3+n}
+ 2^3 \cdot 3^{11} \cdot n^{16} \cdot W \cdot \left(\frac{1}{\alpha}\right)^{28n^3+4n}}_{B^*} \\
& \in O\left( n^{16} \cdot W \cdot \left(\frac{1}{\alpha}\right)^{O(n^3)} \right). \qedhere
\end{align*}
\end{proof}

Using Theorem~\ref{theo:sup-mdp}, for all  $\epsilon > 0$ we can compute a bound $t^*$ such that 
there exists an $\epsilon$-optimal strategy (for expected value) that plays according to an optimal mean-payoff 
strategy after time $t^*$.

\begin{lem}\label{lem:epsilon-optimal-strategy}
Given an MDP~$\M$ and $\epsilon > 0$, there exists an $\epsilon$-optimal strategy 
that plays, after time $t^* = \frac{T\cdot(2B^* + \epsilon)}{\epsilon}$ 
(where $B^*$ is the bound given by Theorem~\ref{theo:sup-mdp}), 
according to a memoryless optimal strategy $\straa_{\MP}$ for the mean-payoff value.
\end{lem}

\begin{proof}
Consider an arbitrary strategy $\straa$ in $\M$ (under expected stopping time $T$),
and given $t^* \geq T$, consider a strategy $\straa^{*}$ that plays like $\straa$ up to time $t^*$,
and then switches to a memoryless mean-payoff optimal strategy $\straa_{\MP}$,
in the MDP $\M$ with initial distribution $\mu^{*} = \delta^{\straa}_{t^*}$ (the 
vertex distribution of $\M$ after $t^*$ steps under strategy $\straa$).
Let $\eta^*$ be the optimal mean-payoff value from $\mu^{*}$ in $\M$,
and let $w' = w - \eta^*$ (where $w'(v) = w(v) - \eta^*$ for all $v \in V$).
With weight vector $w'$, the optimal mean-payoff value of $\M$ is $0$ from $\mu^{*}$.

Using Lemma~\ref{lem:sup-mc} in the Markov chain obtained by fixing the strategy
$\straa_{\MP}$ in $\M$ with initial distribution $\mu^{*}$, we obtain:
\begin{equation}
\sup_{t} \,  \left\lvert \sum_{i=0}^{t} \E^{\straa_{\MP}}_{\mu^{*}}(w'(v_i))  \right\rvert \leq \underbrace{12 \cdot n^6 \cdot W \cdot \bigg(\frac{1}{\alpha}\bigg)^{n^2}}_{C^*}. \label{eq:bound-C}
\end{equation}

Let $u_t = \sum_{i=0}^{t} \E_{\mu}^{\straa}(w(v_i))$ and
let $u^* _t = \sum_{i=0}^{t} \E_{\mu}^{\straa^{*}}(w(v_i))$
be the sequence of expected total reward under strategy $\straa$ and $\straa^{*}$ respectively.
To show $\epsilon$-optimality of $\straa^{*}$, 
take $t^* \geq \frac{T\cdot(2B^* + \epsilon)}{\epsilon}$ 
and show that:
$$ \val(u^* ,T) \geq \val(u,T) - \epsilon$$

The proof is in two steps. First we bound the difference $u_t - u^* _t$ as follows, 
for all $t \geq 1$:

\begin{align*}
 u_t - u^* _t & =  \sum_{i=0}^{t} \E_{\mu}^{\straa}(w(v_i)) - \sum_{i=0}^{t} \E_{\mu}^{\straa^{*}}(w(v_i))    \\
              & =  \sum_{i=0}^{t} \E_{\mu}^{\straa}(w'(v_i)) - \sum_{i=0}^{t} \E_{\mu}^{\straa^{*}}(w'(v_i))  \\
              & \tag*{(since $\E(w(\cdot)) = \E(w'(\cdot)) + \eta^*$)}              \\
              & =  \sum_{i=t^*}^{t} \E_{\mu}^{\straa}(w'(v_i)) - \sum_{i=t^*}^{t} \E_{\mu}^{\straa^{*}}(w'(v_i)) \\ 
              & \tag*{($\straa\text{ and }\straa^{*}$ agree in the first $t^*$ steps)}  \\
              & \leq B^* + C^* \leq  2B^*       \\
              & \tag*{(triangular inequality and bounds}  \\[-3pt]
              & \tag*{given by Theorem~\ref{theo:sup-mdp} and \eqref{eq:bound-C})}  \\
\end{align*}
In a second step, consider an arbitrary bi-Dirac distribution $\delta$ with support $\{t_1,t_2\}$
and expected stopping-time $T$, and consider the difference between the value
of sequences $u_t$ and $u^* _t$ under $\delta$, if $t_2 \geq t^*$ (the difference is $0$ if $t_2 < t^*$):
\begin{align*}
& \E_{\delta}(u) - \E_{\delta}(u^* ) \\
\qquad = & \frac{u_{t_1} (t_2 - T) + u_{t_2} (T - t_1)}{t_2 - t_1} -  \frac{u^* _{t_1} (t_2 - T) + u^* _{t_2} (T - t_1)}{t_2 - t_1}\\
= & \frac{T - t_1}{t_2 - t_1} \cdot (u_{t_2} - u^* _{t_2}) \\
& \tag*{(since $\straa$ and $\straa^{*}$ agree in the first $t^*$ steps,} \\[-3pt]
& \tag*{and thus $u_{t_1} = u^*_{t_1}$)} \\
 \leq & \frac{T - t_1}{t_2 - t_1} \cdot 2B^* \leq \frac{T}{t^* - T} \cdot 2B^* \leq \epsilon   \\
& \tag*{(since $0 \leq t_1 \leq T$)}   \\
\end{align*}

It follows that under all bi-Dirac distributions $\delta$ with expected stopping-time $T$, 
the expected value of the sequence $u^*_t$ is, up to additive error $\epsilon$, greater than 
the expected value of $u_t$. Therefore, since bi-Dirac distributions are sufficient 
for optimality (Section~\ref{sec:Geometric-Interpretation}), we have $\val(u^* ,T) \geq \val(u,T) - \epsilon$.
Hence $\straa^{*}$ is $\epsilon$-optimal.
\end{proof}

We can express in the existential theory of the reals that the value of 
a strategy that eventually plays according to a memoryless 
strategy (as in Lemma~\ref{lem:epsilon-optimal-strategy})
is above a given threshold, which entails decidability of computing an approximation
of the optimal value up to an additive error~$\epsilon$.

\begin{lem}\label{lem:theory-of-reals}
Given an MDP~$\M$ and a time $t^*$, we can compute to an arbitrary level of precision~$\epsilon > 0$ 
the optimal value among the strategies that play after time $t^*$ according to a memoryless strategy.
\end{lem}

\begin{proof}
We describe the choices of an arbitrary strategy up to time $t^*$
using variables $x_{v,t,a}$ for every $v \in V$,
$0\leq t \leq t^*$, and $a \in A$, where $x_{v,t,a}$ is the probability 
to play action $a$ at time $t$ in vertex $v$. Note that we ignore
the history of vertices, which is no loss of generality since the utility achieved
by a strategy at time $t$ only depends on the probability mass in each vertex
at time $t$, and if a sequence of distribution can be achieved by some strategy,
then it can be achieved by a Markov strategy (in which the choice depends only on the
time and the current vertex). We can express the probability mass $p_{v,t}$
in $v$ at time $t$ as $p_{v,t} = \sum_{u \in V} \sum_{a \in A} p_{u,t-1} \cdot x_{u,t-1,a} \cdot \theta(u,a)(v)$ 
where $\theta$ is the transition function of~$\M$.
It is then easy to express the utility $u_t$ as a function of the variables $p_{v,t}$ and $x_{v,t,a}$.

After time $t^*$, consider a memoryless strategy and we can express its mean-payoff
value $\eta^*$ as a function of the vertex distribution at time $t^*$,
thus as a function of the variables $p_{v,t^*}$. Then 
for $t = t^*+1, t^*+2, \dots, \hat{t}$, we express the utility $u_t$ at time $t$
as a function of the variables $x_{v,t,a}$ and $p_{v,t}$, and consider the 
utility sequence $u_0, \dots, u_{\hat{t}}, u_{\hat{t}} + \eta^*, u_{\hat{t}} + 2\eta^*, \dots$ (corresponding to an ultimately periodic path)
using Lemma~\ref{lem:val-cycle} and by an argument similar to the proof of Lemma~\ref{lem:approximation} using 
the bound of Lemma~\ref{lem:sup-mc} for Markov chains, we get a bound on the 
approximation error as follows: the value after $\hat{t}$ differ by at most
$D = n \cdot W \cdot K_1 \cdot K_3^{\hat{t} - t^*}$ from the actual utility, thus the error on the value is at most 
$$\frac{D\cdot  (T-t_1)}{t_2 - t_1} \leq D\cdot T$$
which is at most $\epsilon$ for $\hat{t} \geq t^* + B$ where 
$B = \frac{\ln(\frac{\epsilon}{n \cdot W \cdot T \cdot K_1})}{\ln(K_3)}$ 
(Lemma~\ref{lem:approximation}).
\end{proof}

By Lemma~\ref{lem:epsilon-optimal-strategy} and Lemma~\ref{lem:theory-of-reals},
we can compute up to error $\frac{\epsilon}{2}$ the value of an $\frac{\epsilon}{2}$-optimal strategy,
and since the error is additive ($\epsilon = \frac{\epsilon}{2} +  \frac{\epsilon}{2}$),
it follows from the proof of Lemma~\ref{lem:theory-of-reals} 
that, by computing (as a symbolic expression in variables $x_{v,t,a}$ and $p_{v,t}$) the 
sequence of utilities up to time 
$\hat{t} = \frac{T\cdot(4B^* + \epsilon)}{\epsilon} + \frac{\ln(\frac{\epsilon}{2n \cdot W\cdot T  \cdot K_1})}{\ln(K_3)}$
and then considering an increment of $\eta^*$ at every step, we can compute the value 
of optimal expected value of the MDP up to error $\epsilon$ in exponential space (since $\hat{t}$ is exponential
and the existential theory of the reals can be decided in PSPACE~\cite{Canny88}).
In this way, we obtain the main result of this section: an approximation of the value with 
expected stopping time can be computed for MDPs up to an arbitrary additive error.

\begin{thm}\label{thm:approx-MDP}
The optimal expected value of an MDP with expected stopping time $T$
can be computed to an arbitrary level of precision~$\epsilon > 0$, in exponential space.
\end{thm}

\section{Conclusion}\label{sec:conclusion}
We studied Markov chains and MDPs with expected stopping time, and showed the hardness
of computing the exact value, as the associated decision problem for Markov chains 
is inter-reducible with the Positivity problem, thus at least as hard as the Skolem problem.
Approximation of the value can be computed in exponential time for Markov chains,
and exponential space for MDPs (thus the approximation problem is decidable
although optimal strategies require infinite memory).

\begin{figure}[!tb]
  \begin{center}
    \hrule
      \input{figures/mdp-pure-memory-needed.tex}
    \hrule
      \caption{An MDP where memory is necessary for optimal expected value in pure strategies. \label{fig:mdp-pure-memory-needed}}
  \end{center}
\end{figure}
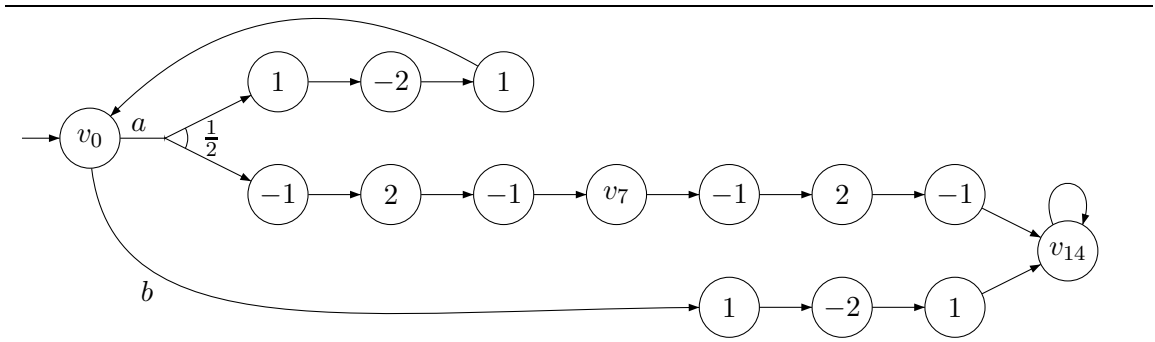

It is an open question to determine the exact complexity of the approximation
problem, and whether approximations can be computed in polynomial time, or 
if any complexity-theoretic lower bound can be established.
We are not aware of any complexity lower bounds for approximation of the Positivity 
problem. Another direction for future work is to determine the memory requirement
for pure strategies in MDPs. \figurename~\ref{fig:mdp-pure-memory-needed} shows 
an MDP where memory is necessary in pure strategies. Consider expected stopping time
$T=8$, and the weight of states $v_0,v_7,v_{14}$ is $0$. A pure strategy that plays
action~$a$ initially in $v_0$ and action~$b$ in the next visit to $v_0$ ensures
expected value of $0$ whereas the expected value of the two memoryless strategies
(playing either always $a$, or always $b$) is negative. This example can be adapted to
show that support-based\footnote{A strategy $\straa$ is support-based
if it plays according to the current state and the support of the current
state distribution. Formally, given a sequence $\rho = v_0 \dots v_k$,
define ${\sf last}(\rho) = v_k$, and for all $i \geq 1$ define 
$S_{\straa}(i) = \{v \in V \mid \exists \rho \in V^i: \P^{\straa}(\rho)>0 \land {\sf last}(\rho) = v\}$.
Then $\straa$ is \emph{support-based} if for all $\rho,\rho'\in V^+$, 
$S_{\straa}(\abs{\rho}) = S_{\straa}(\abs{\rho'})$ and ${\sf last}(\rho) = {\sf last}(\rho')$
imply $\straa(\rho) = \straa(\rho')$.} strategies are not sufficient either.

\medskip\noindent{\bf Acknowledgment.}
The authors are grateful to the anonymous reviewers of LICS 2021 and of a previous
version of this paper for insightful comments that helped improving the 
presentation.
The research presented in this paper was partially supported by the grant ERC CoG 863818 (ForM-SMArt).  

\bibliographystyle{alphaurl}
\bibliography{biblio}

\clearpage
\appendix
\section{Basic Inequalities}

We recall basic inequalities that follow from the
properties of the exponential and logarithm functions.

\begin{lem}\label{lem:math}
For all $x \in \real$:
\begin{enumerate}
\item if $x \geq 1$, then $\big(1-\frac{1}{x}\big)^x < e^{-1} <  \frac{1}{2}$, and
\item if $x < 1$, then $\ln(1-x) \leq -x$. 
\end{enumerate}
\end{lem}


\section{Convergence rate in Markov chains}\label{sec:rate-convergence} 

Let $\tuple{M,\mu,w}$ be an aperiodic Markov chain (i.e., all recurrent classes 
are aperiodic). We show that there exist a vector $\pi$ and
numbers $K_1,K_2$ with $K_2 < 1$ such that 
for all $t \geq 0$ we have: 
$$ \normInf{ \mu \cdot M^t - \pi } \leq K_1 \cdot K_2^t.$$

We recall the following results of~\cite[Chapter~4]{Gallager}.
In every recurrent class (or bottom scc) $C$ of a Markov chain $M$, 
there is a steady-state vector $\pi$ such that 
for all $\mu$ with $\Supp(\mu) \subseteq C$, the vector $\mu \cdot M^t$
converges to $\pi$ as $t \to \infty$.
Moreover by~\cite[Eq.~(4.22)]{Gallager}, for all vertices $i,j \in C$ and for all $t \geq 0$,
we have
\begin{equation}
\abs{M^t_{ij} -  \pi_{j}} \leq \left(1 - 2 \alpha^{n^2}\right)^{\lfloor t/n^2 \rfloor},
\label{eqn:rate-convergence-recurrent}
\end{equation}
where $\alpha = \min \{ M_{ij}  \mid M_{ij >0} \}$ is the smallest non-zero 
probability in $M$, and by~\cite[Lemma~(4.3.6)]{Gallager} for all 
$i \in \T$ where $\T$ is the set of all transient vertices,
\begin{equation}
\sum_{j \in \T} M^t_{ij} \leq (1 - \alpha^{n})^{\lfloor t/n \rfloor}, 
\label{eqn:rate-convergence-transient1}
\end{equation}
which gives a bound on the probability to remain in the set of transient vertices after
$t$ steps. For $i \in V$ and recurrent class $C \subseteq V$, let 
$\P_i(\Diamond C)$ be the probability to eventually reach a vertex in $C$ (and stay there forever
since $C$ is a recurrent class) from $i$. It directly follows from~\eqref{eqn:rate-convergence-transient1} 
that 
\begin{equation}
\P_i(\Diamond C) \geq \sum_{j \in C} M^m_{ij} \geq \P_i(\Diamond C) - (1 - \alpha^{n})^{\lfloor m/n \rfloor}.
\label{eqn:rate-convergence-transient2}
\end{equation}

Now consider a Markov chain~$M$ with only aperiodic recurrent classes, and let $\T$ 
be the set of transient vertices, and $\R$ be the set of recurrent classes. 
The sequence $\mu \cdot M^t$ converges
to a steady-state vector $\pi = \sum_{i \in V} \mu_i \cdot \sum_{C \in \R} \P_i(\Diamond C)\cdot \pi^C$
where $\pi^C$ is the steady-state vector of the class $C$, that is $\pi_j^C = \lim_{t \to \infty} M_{ij}^t$
(for arbitrary $i \in C$, and remember that the limit is independent of $i$).
Let $u = \lfloor t/2 \rfloor$ and $B = \left(1 - 2\alpha^{n^2}\right)^{\lfloor (t-u)/n^2 \rfloor}$. 
Then for all $j \in V$,

\noindent\begin{minipage}{\textwidth}
\begin{align*}
& \left\lvert \sum_{i \in V} \mu_i \cdot M^t_{ij} - \pi_j \right\rvert \\
& = \left\lvert \sum_{i \in V} \mu_i \cdot \left( \sum_{k \in \T} M^u_{ik} M^{t-u}_{kj}
  + \sum_{k \in C}  M^u_{ik} M^{t-u}_{kj}
  - \P_i(\Diamond C)\cdot \pi_j^C \right) \right\rvert \\
\intertext{
(where $C$ is a recurrent class that contains $j$ if $j$ is recurrent, and $C$ is an
arbitrary recurrent class if $j$ is transient, since then $M^{t-u}_{kj} = \pi_j^C  = 0$ for all $k \in C$)
}
& \leq \left\lvert \sum_{i \in V} \mu_i \cdot \left( \sum_{k \in \T} M^u_{ik} 
  + \sum_{k \in C}  M^u_{ik} (\pi_j + B)
  - \P_i(\Diamond C)\cdot \pi_j^C   \right) \right\rvert \\
& \leq \left\lvert \sum_{i \in V} \mu_i \cdot \left(  (1 - \alpha^{n})^{\lfloor u/n \rfloor}
  + B + \sum_{k \in C}  (M^u_{ik}
  - \P_i(\Diamond C)) \cdot \pi_j^C  \right) \right\rvert \\
& \leq (1 - \alpha^{n})^{\lfloor u/n \rfloor} + B + 
  \left\lvert \sum_{i \in V} \mu_i \cdot (1 - \alpha^{n})^{\lfloor u/n \rfloor}  \right\rvert \\
& \leq 2 (1 - \alpha^{n})^{\lfloor u/n \rfloor} + \left(1 - 2\alpha^{n^2}\right)^{\lfloor (t-u)/n^2 \rfloor}  \\
& \leq 3 (1 - \alpha^{n^2})^{t/3n^2}   \\
\end{align*}
\end{minipage}
which proves the result (take $K_1 = 3$ and $K_2 = (1 - \alpha^{n^2})^{1/3n^2}$).

\end{document}

%% file: figures/geometry-intuition-1.tex

\scalebox{0.5}{
\begin{picture}(140,93)(-10,4)

\gasset{Nw=6,Nh=6,Nmr=3, rdist=1, loopdiam=5}

\drawline[AHnb=0,arcradius=1, linegray=.7](52,5)(52,95)

\drawline[AHnb=1,arcradius=1](2,55)(110,55)

\drawline[AHnb=0,arcradius=1](2,53)(2,57)
\drawline[AHnb=0,arcradius=1](52,53)(52,57)

\drawline[AHnb=0,arcradius=1](32,54)(32,56)
\drawline[AHnb=0,arcradius=1](62,54)(62,56)

\node[Nmarks=n, Nframe=n](n1)(2,59){\scalebox{2}{$0$}}
\node[Nmarks=n, Nframe=n](n1)(52,59){\scalebox{2}{$T$}}
\node[Nmarks=n, Nframe=n](n1)(32,59){\scalebox{1.4}{$t_1$}}
\node[Nmarks=n, Nframe=n](n1)(62.3,59){\scalebox{1.4}{$t_2$}}
\node[Nmarks=n, Nframe=n](n1)(110,59){\scalebox{2}{$t$}}

\drawline[AHnb=0,arcradius=1, dash={1.5 2}0 ](2,35)(107,56)
\drawline[AHnb=1,ATnb=1,arcradius=1, linewidth=.35](52,45)(52,55)

\node[Nmarks=n, Nframe=n, Nh=0, Nw=0, Nmr=0](start)(53,50){}
\node[Nmarks=n, Nframe=n, Nh=0, Nw=0, Nmr=0](end)(60,40){}
\node[Nmarks=n, Nframe=n, Nh=0, Nw=0, Nmr=0](label)(90,38){\scalebox{1.4}{optimal value of the sequence}}

\drawqbpedge[AHnb=0](start,340,end,110){}

\node[Nframe=y, Nh=1,Nw=1,Nmr=.5, Nfill=y](n1)(2,55){}
\node[Nframe=y, Nh=1,Nw=1,Nmr=.5, Nfill=y](n2)(7,38){}
\node[Nframe=y, Nh=1,Nw=1,Nmr=.5, Nfill=y](n3)(12,45){}
\node[Nframe=y, Nh=1,Nw=1,Nmr=.5, Nfill=y](n4)(17,42){}
\node[Nframe=y, Nh=1,Nw=1,Nmr=.5, Nfill=y](n5)(22,52){}
\node[Nframe=y, Nh=1,Nw=1,Nmr=.5, Nfill=y](n6)(27,57){}
\node[Nframe=y, Nh=1,Nw=1,Nmr=.5, Nfill=y](n7)(32,41){}
\node[Nframe=y, Nh=1,Nw=1,Nmr=.5, Nfill=y](n8)(37,52){}
\node[Nframe=y, Nh=1,Nw=1,Nmr=.5, Nfill=y](n9)(42,48){}
\node[Nframe=y, Nh=1,Nw=1,Nmr=.5, Nfill=y](n10)(47,70){}
\node[Nframe=y, Nh=1,Nw=1,Nmr=.5, Nfill=y](n11)(52,65){} 
\node[Nframe=y, Nh=1,Nw=1,Nmr=.5, Nfill=y](n12)(57,75){}
\node[Nframe=y, Nh=1,Nw=1,Nmr=.5, Nfill=y](n13)(62,47){} 
\node[Nframe=y, Nh=1,Nw=1,Nmr=.5, Nfill=y](n14)(67,52){}
\node[Nframe=y, Nh=1,Nw=1,Nmr=.5, Nfill=y](n15)(72,65){}
\node[Nframe=y, Nh=1,Nw=1,Nmr=.5, Nfill=y](n16)(77,62){}
\node[Nframe=y, Nh=1,Nw=1,Nmr=.5, Nfill=y](n17)(82,75){}
\node[Nframe=y, Nh=1,Nw=1,Nmr=.5, Nfill=y](n18)(87,72){}
\node[Nframe=y, Nh=1,Nw=1,Nmr=.5, Nfill=y](n19)(92,85){}
\node[Nframe=y, Nh=1,Nw=1,Nmr=.5, Nfill=y](n20)(97,82){}
\node[Nframe=n, Nh=1,Nw=1,Nmr=.5](n21)(99,86){}


\drawedge[ELpos=50, ELside=r, curvedepth=0, AHnb=0, linegray=.5](n1,n2){}
\drawedge[ELpos=50, ELside=r, curvedepth=0, AHnb=0, linegray=.5](n2,n3){}
\drawedge[ELpos=50, ELside=r, curvedepth=0, AHnb=0, linegray=.5](n3,n4){}
\drawedge[ELpos=50, ELside=r, curvedepth=0, AHnb=0, linegray=.5](n4,n5){}
\drawedge[ELpos=50, ELside=r, curvedepth=0, AHnb=0, linegray=.5](n5,n6){}
\drawedge[ELpos=50, ELside=r, curvedepth=0, AHnb=0, linegray=.5](n6,n7){}
\drawedge[ELpos=50, ELside=r, curvedepth=0, AHnb=0, linegray=.5](n7,n8){}
\drawedge[ELpos=50, ELside=r, curvedepth=0, AHnb=0, linegray=.5](n8,n9){}
\drawedge[ELpos=50, ELside=r, curvedepth=0, AHnb=0, linegray=.5](n9,n10){}
\drawedge[ELpos=50, ELside=r, curvedepth=0, AHnb=0, linegray=.5](n10,n11){}
\drawedge[ELpos=50, ELside=r, curvedepth=0, AHnb=0, linegray=.5](n11,n12){}
\drawedge[ELpos=50, ELside=r, curvedepth=0, AHnb=0, linegray=.5](n12,n13){}
\drawedge[ELpos=50, ELside=r, curvedepth=0, AHnb=0, linegray=.5](n13,n14){}
\drawedge[ELpos=50, ELside=r, curvedepth=0, AHnb=0, linegray=.5](n14,n15){}
\drawedge[ELpos=50, ELside=r, curvedepth=0, AHnb=0, linegray=.5](n15,n16){}
\drawedge[ELpos=50, ELside=r, curvedepth=0, AHnb=0, linegray=.5](n16,n17){}
\drawedge[ELpos=50, ELside=r, curvedepth=0, AHnb=0, linegray=.5](n17,n18){}
\drawedge[ELpos=50, ELside=r, curvedepth=0, AHnb=0, linegray=.5](n18,n19){}
\drawedge[ELpos=50, ELside=r, curvedepth=0, AHnb=0, linegray=.5](n19,n20){}
\drawedge[ELpos=50, ELside=r, curvedepth=0, AHnb=0, linegray=.5, dash={1}0](n20,n21){}








\end{picture}
}

%% file: figures/geometry-intuition-2.tex

\scalebox{0.5}{
\begin{picture}(140,93)(-10,4)

\gasset{Nw=6,Nh=6,Nmr=3, rdist=1, loopdiam=5}

\drawline[AHnb=0,arcradius=1, linegray=.7](52,5)(52,95)

\drawline[AHnb=1,arcradius=1](2,55)(110,55)

\drawline[AHnb=0,arcradius=1](2,53)(2,57)
\drawline[AHnb=0,arcradius=1](52,53)(52,57)
\drawline[AHnb=0,arcradius=1](42,54)(42,56)

\node[Nmarks=n, Nframe=n](n1)(2,59){\scalebox{2}{$0$}}
\node[Nmarks=n, Nframe=n](n1)(52,59){\scalebox{2}{$T$}}
\node[Nmarks=n, Nframe=n](n1)(42,59){\scalebox{1.4}{$t_1$}}
\node[Nmarks=n, Nframe=n](n1)(110,59){\scalebox{2}{$t$}}

\drawline[AHnb=0,arcradius=1, dash={1.5 2}0 ](22,5)(97,80)
\drawline[AHnb=1,ATnb=1,arcradius=1, linewidth=.45](52,35)(52,55)

\node[Nmarks=n, Nframe=n, Nh=0, Nw=0, Nmr=0](start)(53,45){}
\node[Nmarks=n, Nframe=n, Nh=0, Nw=0, Nmr=0](end)(60,35){}
\node[Nmarks=n, Nframe=n, Nh=0, Nw=0, Nmr=0](label)(90,33){\scalebox{1.4}{optimal value of the sequence}}

\drawqbpedge[AHnb=0](start,340,end,110){}

\node[Nframe=y, Nh=1,Nw=1,Nmr=.5, Nfill=y](n1)(2,55){}
\node[Nframe=y, Nh=1,Nw=1,Nmr=.5, Nfill=y](n2)(7,40){}
\node[Nframe=y, Nh=1,Nw=1,Nmr=.5, Nfill=y](n3)(12,45){}
\node[Nframe=y, Nh=1,Nw=1,Nmr=.5, Nfill=y](n4)(17,38){}
\node[Nframe=y, Nh=1,Nw=1,Nmr=.5, Nfill=y](n5)(22,15){}
\node[Nframe=y, Nh=1,Nw=1,Nmr=.5, Nfill=y](n6)(27,60){}
\node[Nframe=y, Nh=1,Nw=1,Nmr=.5, Nfill=y](n7)(32,70){}
\node[Nframe=y, Nh=1,Nw=1,Nmr=.5, Nfill=y](n8)(37,52){}
\node[Nframe=y, Nh=1,Nw=1,Nmr=.5, Nfill=y](n9)(42,25){} 
\node[Nframe=y, Nh=1,Nw=1,Nmr=.5, Nfill=y](n10)(47,70){}
\node[Nframe=y, Nh=1,Nw=1,Nmr=.5, Nfill=y](n11)(52,65){} 
\node[Nframe=y, Nh=1,Nw=1,Nmr=.5, Nfill=y](n12)(57,75){}
\node[Nframe=y, Nh=1,Nw=1,Nmr=.5, Nfill=y](n13)(62,60){}
\node[Nframe=y, Nh=1,Nw=1,Nmr=.5, Nfill=y](n14)(67,75){}
\node[Nframe=y, Nh=1,Nw=1,Nmr=.5, Nfill=y](n15)(72,70){}
\node[Nframe=y, Nh=1,Nw=1,Nmr=.5, Nfill=y](n16)(77,85){}
\node[Nframe=y, Nh=1,Nw=1,Nmr=.5, Nfill=y](n17)(82,80){}
\node[Nframe=y, Nh=1,Nw=1,Nmr=.5, Nfill=y](n18)(87,95){}
\node[Nframe=y, Nh=1,Nw=1,Nmr=.5, Nfill=y](n19)(92,90){}
\node[Nframe=n, Nh=1,Nw=1,Nmr=.5](n20)(94,96){}

\drawedge[ELpos=50, ELside=r, curvedepth=0, AHnb=0, linegray=.5](n1,n2){}
\drawedge[ELpos=50, ELside=r, curvedepth=0, AHnb=0, linegray=.5](n2,n3){}
\drawedge[ELpos=50, ELside=r, curvedepth=0, AHnb=0, linegray=.5](n3,n4){}
\drawedge[ELpos=50, ELside=r, curvedepth=0, AHnb=0, linegray=.5](n4,n5){}
\drawedge[ELpos=50, ELside=r, curvedepth=0, AHnb=0, linegray=.5](n5,n6){}
\drawedge[ELpos=50, ELside=r, curvedepth=0, AHnb=0, linegray=.5](n6,n7){}
\drawedge[ELpos=50, ELside=r, curvedepth=0, AHnb=0, linegray=.5](n7,n8){}
\drawedge[ELpos=50, ELside=r, curvedepth=0, AHnb=0, linegray=.5](n8,n9){}
\drawedge[ELpos=50, ELside=r, curvedepth=0, AHnb=0, linegray=.5](n9,n10){}
\drawedge[ELpos=50, ELside=r, curvedepth=0, AHnb=0, linegray=.5](n10,n11){}
\drawedge[ELpos=50, ELside=r, curvedepth=0, AHnb=0, linegray=.5](n11,n12){}
\drawedge[ELpos=50, ELside=r, curvedepth=0, AHnb=0, linegray=.5](n12,n13){}
\drawedge[ELpos=50, ELside=r, curvedepth=0, AHnb=0, linegray=.5](n13,n14){}
\drawedge[ELpos=50, ELside=r, curvedepth=0, AHnb=0, linegray=.5](n14,n15){}
\drawedge[ELpos=50, ELside=r, curvedepth=0, AHnb=0, linegray=.5](n15,n16){}
\drawedge[ELpos=50, ELside=r, curvedepth=0, AHnb=0, linegray=.5](n16,n17){}
\drawedge[ELpos=50, ELside=r, curvedepth=0, AHnb=0, linegray=.5](n17,n18){}
\drawedge[ELpos=50, ELside=r, curvedepth=0, AHnb=0, linegray=.5](n18,n19){}
\drawedge[ELpos=50, ELside=r, curvedepth=0, AHnb=0, linegray=.5, dash={1}0](n19,n20){}








\end{picture}
}

%% file: figures/reduction-chains-compact.tex

\begin{picture}(135,56)(0,0)

\gasset{Nw=6,Nh=8,Nmr=1.5, rdist=1, loopdiam=5}

\node[Nframe=y, Nw=20](n1)(10,40){\phantom{${}^{>}$}Positivity\phantom{${}^{>}$}}
\node[Nframe=y, Nw=24](n2)(55,40){\phantom{${}^{>}$}Problem A${}^{>}$\phantom{${}^{>}$}}
\node[Nframe=y, Nw=40](n3)(115,40){\phantom{${}^{>}$}Markov Reachability${}^{>}$\phantom{${}^{>}$}}

\node[Nframe=y, Nw=50, Nh=12](n4)(85,10){\begin{tabular}{c}Exact value problem \\ with expected stopping time \end{tabular}}

\drawedge[ELpos=46, ELside=l, curvedepth=0](n1,n2){\cite{AAOW15}}
\drawedge[ELpos=40, ELside=l, curvedepth=0](n2,n3){\cite{AAOW15}}

\drawedge[ELpos=50, ELside=r, syo=4, eyo=4, dash={1}0, curvedepth=-6](n3,n1){Lemma~\ref{lem:markov-to-pos}}

\drawedge[ELpos=25, ELside=r, dash={1}0, curvedepth=-6](n2,n4){Lemma~\ref{lem:pos-to-exact}}
\drawedge[ELpos=35, ELside=r, dash={1}0, sxo=-4, curvedepth=-6](n4,n2){Lemma~\ref{lem:exact-to-A}}






\end{picture}

%% file: figures/Positivity-reduction.tex

\scalebox{0.5}{
\begin{picture}(190,85)(-4,10)

\gasset{Nw=6,Nh=6,Nmr=3, rdist=1, loopdiam=5}

\drawline[AHnb=1,arcradius=1](2,55)(135,55)

\drawline[AHnb=0,arcradius=1, linegray=.7](7,20)(7,75)

\drawline[AHnb=0,arcradius=1](2,53)(2,57)
\drawline[AHnb=0,arcradius=1](7,53)(7,57)

\node[Nmarks=n, Nframe=n](n1)(2,61){\scalebox{2}{$0$}}
\node[Nmarks=n, Nframe=n](n1)(8,50){\scalebox{2}{$T$}}
\node[Nmarks=n, Nframe=n](n1)(135,59){\scalebox{2}{$t$}}

\node[Nmarks=n, Nframe=n](n1)(0,25){\scalebox{2}{$a$}}


\drawline[AHnb=0,arcradius=1, dash={1.5 2}0](0,30)(122,30)      


\node[Nmarks=n, Nframe=n](n1)(7,84){\makebox(0,0)[l]{\scalebox{1.6}{\begin{tabular}{l}start simulating \\Markov chain $M$ \end{tabular}}}}
\drawline[arcradius=1](7,85)(7,70)

\node[Nmarks=n, Nframe=n](n1)(120,35){\makebox(0,0)[l]{\scalebox{1.6}{\begin{tabular}{l}asymptotic \\reward\end{tabular}}}}
\drawline[arcradius=1](110,28)(125,28)

\node[Nmarks=n, Nframe=n](n1)(18,22){\makebox(0,0)[l]{\scalebox{1.6}{optimal value if total reward always remains above asymptote}}}
\drawline[arcradius=1](16,22)(9,29)


\node[Nframe=y, Nh=1,Nw=1,Nmr=.5, Nfill=y, fillgray=.5 ,linegray=.5](n0)(2,55){}
\node[Nframe=y, Nh=1,Nw=1,Nmr=.5, Nfill=y, fillgray=.5 ,linegray=.5](n1)(2,30){}
\node[Nframe=y, Nh=1,Nw=1,Nmr=.5, Nfill=y, fillgray=.5 ,linegray=.5](n2)(7,67){}
\node[Nframe=y, Nh=1,Nw=1,Nmr=.5, Nfill=y, fillgray=.5 ,linegray=.5](n3)(12,58){}
\node[Nframe=y, Nh=1,Nw=1,Nmr=.5, Nfill=y, fillgray=.5 ,linegray=.5](n4)(17,62){}
\node[Nframe=y, Nh=1,Nw=1,Nmr=.5, Nfill=y, fillgray=.5 ,linegray=.5](n5)(22,48.75){}
\node[Nframe=y, Nh=1,Nw=1,Nmr=.5, Nfill=y, fillgray=.5 ,linegray=.5](n6)(27,53){}
\node[Nframe=y, Nh=1,Nw=1,Nmr=.5, Nfill=y, fillgray=.5 ,linegray=.5](n7)(32,65){}
\node[Nframe=y, Nh=1,Nw=1,Nmr=.5, Nfill=y, fillgray=.5 ,linegray=.5](n8)(37,62){}
\node[Nframe=y, Nh=1,Nw=1,Nmr=.5, Nfill=y, fillgray=.5 ,linegray=.5](n9)(42,60){}
\node[Nframe=y, Nh=1,Nw=1,Nmr=.5, Nfill=y, fillgray=.5 ,linegray=.5](n10)(47,45){}
\node[Nframe=y, Nh=1,Nw=1,Nmr=.5, Nfill=y, fillgray=.5 ,linegray=.5](n11)(52,56){}
\node[Nframe=y, Nh=1,Nw=1,Nmr=.5, Nfill=y, fillgray=.5 ,linegray=.5](n12)(57,57){}
\node[Nframe=y, Nh=1,Nw=1,Nmr=.5, Nfill=y, fillgray=.5 ,linegray=.5](n13)(62,62){} 
\node[Nframe=y, Nh=1,Nw=1,Nmr=.5, Nfill=y, fillgray=.5 ,linegray=.5](n14)(67,43){}
\node[Nframe=y, Nh=1,Nw=1,Nmr=.5, Nfill=y, fillgray=.5 ,linegray=.5](n15)(72,35){}
\node[Nframe=y, Nh=1,Nw=1,Nmr=.5, Nfill=y, fillgray=.5 ,linegray=.5](n16)(77,39){}
\node[Nframe=y, Nh=1,Nw=1,Nmr=.5, Nfill=y, fillgray=.5 ,linegray=.5](n17)(82,34){}
\node[Nframe=y, Nh=1,Nw=1,Nmr=.5, Nfill=y, fillgray=.5 ,linegray=.5](n18)(87,33){}
\node[Nframe=y, Nh=1,Nw=1,Nmr=.5, Nfill=y, fillgray=.5 ,linegray=.5](n19)(92,36){} 
\node[Nframe=y, Nh=1,Nw=1,Nmr=.5, Nfill=y, fillgray=.5 ,linegray=.5](n20)(97,31){}
\node[Nframe=y, Nh=1,Nw=1,Nmr=.5, Nfill=y, fillgray=.5 ,linegray=.5](n21)(102,32){}
\node[Nframe=y, Nh=1,Nw=1,Nmr=.5, Nfill=y, fillgray=.5 ,linegray=.5](n22)(107,34){}
\node[Nframe=y, Nh=1,Nw=1,Nmr=.5, Nfill=y, fillgray=.5 ,linegray=.5](n23)(112,31){}
\node[Nframe=y, Nh=1,Nw=1,Nmr=.5, Nfill=y, fillgray=.5 ,linegray=.5](n24)(117,33){}

\gasset{linegray=.5}
\drawedge[ELpos=50, ELside=r, curvedepth=0, AHnb=0](n0,n1){}
\drawedge[ELpos=50, ELside=r, curvedepth=0, AHnb=0](n1,n2){}
\drawedge[ELpos=50, ELside=r, curvedepth=0, AHnb=0](n2,n3){}
\drawedge[ELpos=50, ELside=r, curvedepth=0, AHnb=0](n3,n4){}
\drawedge[ELpos=50, ELside=r, curvedepth=0, AHnb=0](n4,n5){}
\drawedge[ELpos=50, ELside=r, curvedepth=0, AHnb=0](n5,n6){}
\drawedge[ELpos=50, ELside=r, curvedepth=0, AHnb=0](n6,n7){}
\drawedge[ELpos=50, ELside=r, curvedepth=0, AHnb=0](n7,n8){}
\drawedge[ELpos=50, ELside=r, curvedepth=0, AHnb=0](n8,n9){}
\drawedge[ELpos=50, ELside=r, curvedepth=0, AHnb=0](n9,n10){}
\drawedge[ELpos=50, ELside=r, curvedepth=0, AHnb=0](n10,n11){}
\drawedge[ELpos=50, ELside=r, curvedepth=0, AHnb=0](n11,n12){}
\drawedge[ELpos=50, ELside=r, curvedepth=0, AHnb=0](n12,n13){}
\drawedge[ELpos=50, ELside=r, curvedepth=0, AHnb=0](n13,n14){}
\drawedge[ELpos=50, ELside=r, curvedepth=0, AHnb=0](n14,n15){}
\drawedge[ELpos=50, ELside=r, curvedepth=0, AHnb=0](n15,n16){}
\drawedge[ELpos=50, ELside=r, curvedepth=0, AHnb=0](n16,n17){}
\drawedge[ELpos=50, ELside=r, curvedepth=0, AHnb=0](n17,n18){}
\drawedge[ELpos=50, ELside=r, curvedepth=0, AHnb=0](n18,n19){}
\drawedge[ELpos=50, ELside=r, curvedepth=0, AHnb=0](n19,n20){}
\drawedge[ELpos=50, ELside=r, curvedepth=0, AHnb=0](n20,n21){}
\drawedge[ELpos=50, ELside=r, curvedepth=0, AHnb=0](n21,n22){}
\drawedge[ELpos=50, ELside=r, curvedepth=0, AHnb=0](n22,n23){}
\drawedge[ELpos=50, ELside=r, curvedepth=0, AHnb=0](n23,n24){}






\end{picture}
}

%% file: figures/Positivity-oracle.tex

\scalebox{0.5}{
\begin{picture}(185,92)(-10,26)

\gasset{Nw=6,Nh=6,Nmr=3, rdist=1, loopdiam=5}

\drawline[AHnb=1,arcradius=1](2,75)(130,75)

\drawline[AHnb=0,arcradius=1, linegray=.7](52,30)(52,115)

\drawline[AHnb=0,arcradius=1](2,73)(2,77)
\drawline[AHnb=0,arcradius=1](52,73)(52,77)

\drawline[AHnb=0,arcradius=1](50,55)(54,55)
\node[Nmarks=n, Nframe=y, Nh=1,Nw=1,Nmr=.5, Nfill=y, fillcolor=red, linecolor=red](n1)(52,55){}

\node[Nmarks=n, Nframe=n](n1)(2,69){\scalebox{2}{$0$}}
\node[Nmarks=n, Nframe=n](n1)(52,69){\scalebox{2}{$T$}}
\node[Nmarks=n, Nframe=n](n1)(130,79){\scalebox{2}{$t$}}
\node[Nmarks=n, Nframe=n](n1)(57,54){\scalebox{2}{$\nu$}}

\drawline[AHnb=0,arcradius=1, linecolor=red](2,30)(122,90)      


\drawline[AHnb=0,arcradius=1, dash={1 1.5}0, linegray=.4](57,100)(62,105)(67,99)(72,92)(77,105)(82,112)(87,95)(92,89)(97,92)(102,89)(107,95)(112,103)(117,96)(122,105)
\drawline[AHnb=0,arcradius=1, dash={1 1.5}0, linegray=.4](57,94)(62,90)(67,96)(72,85)(77,79)(82,82)(87,91)(92,76)(97,80)(102,85)(107,88)(112,87)(117,90)(122,91)
\drawline[AHnb=0,arcradius=1, dash={1 1.5}0, linegray=.4](57,85)(62,80)(67,78)(72,81)(77,72)(82,65)(87,66)(92,60)(97,61)(102,63)(107,72)(112,68)(117,71)(122,78)


\node[Nmarks=n, Nframe=n](n1)(60,40){\makebox(0,0)[l]{\scalebox{1.6}{\begin{tabular}{l}value is at least $\nu$ if the sequence of utilities\\ 
remains above the bottom line\end{tabular}}}}
\drawline[arcradius=1](72,48)(65,60) 
\node[Nmarks=n, Nframe=n](n1)(64,87){\makebox(0,0)[l]{\scalebox{2}?}}


\node[Nframe=y, Nh=1,Nw=1,Nmr=.5, Nfill=y, fillgray=.5 ,linegray=.5](n1)(2,75){}
\node[Nframe=y, Nh=1,Nw=1,Nmr=.5, Nfill=y, fillgray=.5 ,linegray=.5](n2)(7,77){}
\node[Nframe=y, Nh=1,Nw=1,Nmr=.5, Nfill=y, fillgray=.5 ,linegray=.5](n3)(12,84){}
\node[Nframe=y, Nh=1,Nw=1,Nmr=.5, Nfill=y, fillgray=.5 ,linegray=.5](n4)(17,82){}
\node[Nframe=y, Nh=1,Nw=1,Nmr=.5, Nfill=y, fillgray=.5 ,linegray=.5](n5)(22,48){}
\node[Nframe=y, Nh=1,Nw=1,Nmr=.5, Nfill=y, fillgray=.5 ,linegray=.5](n6)(27,53){}
\node[Nframe=y, Nh=1,Nw=1,Nmr=.5, Nfill=y, fillgray=.5 ,linegray=.5](n7)(32,45){}
\node[Nframe=y, Nh=1,Nw=1,Nmr=.5, Nfill=y, fillgray=.5 ,linegray=.5](n8)(37,63){}
\node[Nframe=y, Nh=1,Nw=1,Nmr=.5, Nfill=y, fillgray=.5 ,linegray=.5](n9)(42,58){}
\node[Nframe=y, Nh=1,Nw=1,Nmr=.5, Nfill=y, fillgray=.5 ,linegray=.5](n10)(47,82){}
\node[Nframe=y, Nh=1,Nw=1,Nmr=.5, Nfill=y, fillgray=.5 ,linegray=.5](n11)(52,90){}
\node[Nframe=y, Nh=1,Nw=1,Nmr=.5, Nfill=y, fillgray=.5 ,linegray=.5](n12a)(57,100){}
\node[Nframe=y, Nh=1,Nw=1,Nmr=.5, Nfill=y, fillgray=.5 ,linegray=.5](n12b)(57,94){}
\node[Nframe=y, Nh=1,Nw=1,Nmr=.5, Nfill=y, fillgray=.5 ,linegray=.5](n12c)(57,85){}




\gasset{linegray=.5}
\drawedge[ELpos=50, ELside=r, curvedepth=0, AHnb=0](n1,n2){}
\drawedge[ELpos=50, ELside=r, curvedepth=0, AHnb=0](n2,n3){}
\drawedge[ELpos=50, ELside=r, curvedepth=0, AHnb=0](n3,n4){}
\drawedge[ELpos=50, ELside=r, curvedepth=0, AHnb=0](n4,n5){}
\drawedge[ELpos=50, ELside=r, curvedepth=0, AHnb=0](n5,n6){}
\drawedge[ELpos=50, ELside=r, curvedepth=0, AHnb=0](n6,n7){}
\drawedge[ELpos=50, ELside=r, curvedepth=0, AHnb=0](n7,n8){}
\drawedge[ELpos=50, ELside=r, curvedepth=0, AHnb=0](n8,n9){}
\drawedge[ELpos=50, ELside=r, curvedepth=0, AHnb=0](n9,n10){}
\drawedge[ELpos=50, ELside=r, curvedepth=0, AHnb=0](n10,n11){}
\drawedge[ELpos=50, ELside=r, curvedepth=0, AHnb=0, dash={1 1.5}0, linegray=.4](n11,n12a){}
\drawedge[ELpos=50, ELside=r, curvedepth=0, AHnb=0, dash={1 1.5}0, linegray=.4](n11,n12b){}
\drawedge[ELpos=50, ELside=r, curvedepth=0, AHnb=0, dash={1 1.5}0, linegray=.4](n11,n12c){}







\end{picture}
}

%% file: figures/markov-chain-lower-bound.tex
\def\smallestprob{$\alpha$}
\def\cosmallestprob{$1\!-\!\alpha$}

\begin{picture}(150,45)(0,2)

\gasset{Nw=8,Nh=8,Nmr=4, rdist=1, loopdiam=5}

\node[Nframe=y, Nmarks=i](n0)(10,35){$0$}
\node[Nframe=y](n1)(34,35){$1$}
\node[Nframe=y](n2)(58,35){$2$}
\node[Nframe=n, Nw=0, Nh=0](n3s)(72,35){}
\node[Nframe=n, Nw=0, Nh=0](n3dots)(77,35){$\dots$}
\node[Nframe=n, Nw=0, Nh=0](n3e)(82,35){}
\node[Nframe=y](nn2)(97,35){{\scriptsize $n\!-\!2$}}
\node[Nframe=y](nn1)(121,35){{\scriptsize $n\!-\!1$}}
\node[Nframe=y](nn)(145,35){$n$}

\node[Nframe=n, Nw=0, Nh=0](a3s)(68,27){}
\node[Nframe=n, Nw=0, Nh=0](a3dots)(71,25.8){\rotatebox[origin=c]{-4}{$\ddots$}}

\node[Nframe=n, Nw=0, Nh=0](b3dots)(84.2,26.5){\rotatebox[origin=c]{-4}{$\ddots$}}
\node[Nframe=n, Nw=0, Nh=0](b3e)(87,23){}

\node[Nframe=y](nb1)(34,15){$\bar{1}$}
\node[Nframe=y](nb2)(58,15){$\bar{2}$}
\node[Nframe=n, Nw=0, Nh=0](nb3s)(72,15){}
\node[Nframe=n, Nw=0, Nh=0](nb3dots)(77,15){$\dots$}
\node[Nframe=n, Nw=0, Nh=0](nb3e)(82,15){}
\node[Nframe=y](nbn2)(97,15){{\scriptsize $\overline{n\!-\!2}$}}
\node[Nframe=y](nbn1)(121,15){{\scriptsize $\overline{n\!-\!1}$}}


\drawedge[ELpos=30, ELside=l, curvedepth=0](n0,n1){\smallestprob}
\drawedge[ELpos=30, ELside=l, curvedepth=0](n1,n2){\smallestprob}
\drawedge[ELpos=50, ELside=l, curvedepth=0](n2,n3s){\smallestprob}
\drawedge[ELpos=60, ELside=l, curvedepth=0](n3e,nn2){}
\drawedge[ELpos=30, ELside=l, curvedepth=0](nn2,nn1){\smallestprob}
\drawedge[ELpos=30, ELside=l, curvedepth=0](nn1,nn){\smallestprob}
\drawbpedge[ELpos=12, ELside=r, syo=3, eyo=3, ELdist=0.5, curvedepth=6](nn1,135,10,n0,75,10){\cosmallestprob}
\drawloop[ELside=l, loopCW=y, loopdiam=6](nn){}
\drawedge[ELpos=28, ELside=r, ELdist=-2.5, curvedepth=0](n0,nb1){\rotatebox[origin=c]{-38.6}{\cosmallestprob}}
\drawedge[ELpos=28, ELside=r, ELdist=-2.5, curvedepth=0](n1,nb2){\rotatebox[origin=c]{-38.6}{\cosmallestprob}}
\drawedge[ELpos=28, ELside=r, ELdist=-2.5, curvedepth=0](nn2,nbn1){\rotatebox[origin=c]{-38.6}{\cosmallestprob}}

\drawedge[ELpos=70, ELside=r, ELdist=-2.5, curvedepth=0](n2,a3s){\rotatebox[origin=c]{-38.6}{\cosmallestprob}}
\drawedge[ELpos=50, ELside=r, curvedepth=0](b3e,nbn2){}

\drawedge[ELpos=30, ELside=r, curvedepth=0](nb1,nb2){}
\drawedge[ELpos=60, ELside=r, curvedepth=0](nb2,nb3s){}
\drawedge[ELpos=60, ELside=l, curvedepth=0](nb3e,nbn2){}
\drawedge[ELpos=60, ELside=l, curvedepth=0](nbn2,nbn1){}
\drawqbpedge[ELpos=60, ELside=l, syo=-3, curvedepth=6](nbn1,190,n0,265){}






\end{picture}

%% file: figures/mdp-infinite-mem.tex
\begin{picture}(80,86)(-5,0)

\gasset{Nw=8,Nh=8,Nmr=4, rdist=1, loopdiam=5}

\node[Nframe=y, Nmarks=i](n0)(5,43){$v_0$}

\node[Nframe=y, Nmarks=n](m4)(25,63){$v_1$}
\node[Nframe=y, Nmarks=n](m5)(45,70){$v_2$}
\node[Nframe=y, Nmarks=n](m6)(65,77){$v_3$}
\node[Nframe=y, Nmarks=n](m7)(45,56){$v_4$}
\nodelabel[ExtNL=y, NLangle=270, NLdist=1](m7){$-1$}
\node[Nframe=y, Nmarks=n](m8)(65,63){$v_5$}
\nodelabel[ExtNL=y, NLangle=0, NLdist=1](m8){$2$}
\node[Nframe=y, Nmarks=n](m9)(65,49){$v_6$}
\nodelabel[ExtNL=y, NLangle=0, NLdist=1](m9){$-1$}

\node[Nframe=y, Nmarks=n](n4)(25,23){$v'_1$}
\node[Nframe=y, Nmarks=n](n5)(45,30){$v'_2$}
\node[Nframe=y, Nmarks=n](n6)(65,37){$v'_3$}
\node[Nframe=y, Nmarks=n](n7)(45,16){$v'_4$}
\nodelabel[ExtNL=y, NLangle=270, NLdist=1](n7){$1$}
\node[Nframe=y, Nmarks=n](n8)(65,23){$v'_5$}
\nodelabel[ExtNL=y, NLangle=0, NLdist=1](n8){$-2$}
\node[Nframe=y, Nmarks=n](n9)(65,9){$v'_6$}
\nodelabel[ExtNL=y, NLangle=0, NLdist=1](n9){$1$}






\drawedge[ELpos=31, ELside=r, ELdist=0.2, curvedepth=0](n0,m4){$\frac{1}{3}$}
\drawedge[ELpos=31, ELside=l, ELdist=0.2, curvedepth=0](n0,n4){$\frac{2}{3}$}
\drawarc(5,43,7,315,45)

\drawedge[ELpos=50, ELside=l, curvedepth=0](m4,m5){}
\drawarc(25,63,7,340.71,19.29)
\nodelabel[ExtNL=y, NLangle=0, NLdist=4](m4){$\frac{1}{2}$}
\drawedge[ELpos=50, ELside=l, curvedepth=0](m5,m6){}
\drawedge[ELpos=50, ELside=r, curvedepth=-8](m6,m4){}

\drawedge[ELpos=50, ELside=l, curvedepth=0](m4,m7){}
\drawedge[ELpos=50, ELside=l, curvedepth=0](m7,m8){}      
\drawedge[ELpos=50, ELside=l, curvedepth=0](m8,m9){}      
\drawedge[ELpos=50, ELside=l, curvedepth=0](m9,m7){}      

\drawedge[ELpos=50, ELside=l, curvedepth=0](n4,n5){}
\drawedge[ELpos=50, ELside=l, curvedepth=0](n5,n6){}
\drawedge[ELpos=50, ELside=r, curvedepth=-8](n6,n4){}

\drawedge[ELpos=50, ELside=l, curvedepth=0](n4,n7){}
\drawedge[ELpos=50, ELside=l, curvedepth=0](n7,n8){}      
\drawedge[ELpos=50, ELside=l, curvedepth=0](n8,n9){}      
\drawedge[ELpos=50, ELside=l, curvedepth=0](n9,n7){}      








\end{picture}

%% file: figures/proof-sequence.tex

\begin{picture}(150,58)(0,0)

\gasset{Nw=8,Nh=8,Nmr=4, rdist=1, loopdiam=5}

\node[Nframe=y, Nmarks=n, Nh=22, Nw=28, Nmr=0](l1)(30,44){
\!\!\!\begin{tabular}{l}Markov chains \\[+2pt] $\val^{\MP}(\cdot)\! = 0$ \\[+4pt] 
Lemma~\ref{lem:sup-mc} \\ \phantom{solution} \end{tabular}}

\node[Nframe=y, Nmarks=n, Nh=22, Nw=32, Nmr=0](l2)(70,44){
\!\!\!\begin{tabular}{l}MDP, single EC \\[+2pt] with $\val^{\MP}(\cdot)\! \leq 0$ \\[+4pt] 
Lemma~\ref{lem:sup-mdp-ec} \\ reduction to MC \end{tabular}}

\drawline[AHnb=1,arcradius=1.5](54,36)(49,36)(49,51.5)(44,51.5)

\node[Nframe=y, Nmarks=n, Nh=22, Nw=38, Nmr=0](l3)(115,44){
\!\!\!\begin{tabular}{l}MDP, all EC \\[+2pt] have $\val^{\MP}(\cdot)\! \leq 0$ \\[+4pt] 
Lemma~\ref{lem:sup-mdp-several-ec} \\ back-edge transform. \end{tabular}}

\drawline[AHnb=1,arcradius=1.5](96,36)(91,36)(91,51.5)(86,51.5)

\node[Nframe=y, Nmarks=n, Nh=22, Nw=30, Nmr=0](l5)(53,14){
\!\!\!\begin{tabular}{l}Arbitrary MDP \\[+2pt] $\val^{\MP}(\cdot)\! \leq 0$ \\[+4pt] 
Lemma~\ref{lem:sup-mdp-transform} \\ uniformization \end{tabular}}

\node[Nframe=y, Nmarks=n, Nh=22, Nw=44, Nmr=0](l4)(100,14){
\!\!\!\begin{tabular}{l}Uniform MDP \\[+2pt] $\val^{\MP}(\cdot)\! \leq 0$ \\[+4pt] 
Lemma~\ref{lem:sup-mdp-ec-uniform} \\ reduction to all EC $\leq 0$ \end{tabular}}

\drawline[AHnb=1,arcradius=1.5](68,6)(73,6)(73,21.5)(78,21.5)

\drawline[AHnb=1,arcradius=1.5](122,6)(139,6)(139,51.5)(134,51.5)

%












\end{picture}

%% file: figures/mdp-reset.tex

\begin{picture}(60,35)(0,2)

\gasset{Nw=8,Nh=8,Nmr=4, rdist=1, loopdiam=5}

\node[Nframe=y, Nmarks=n, ExtNL=y, NLdist=1, Nh=6, Nw=6, Nmr=3](n0)(11,20){$v_0$}
\node[Nframe=y, Nmarks=n, ExtNL=y, NLdist=1, Nh=6, Nw=6, Nmr=3](n1)(19,24){}
\node[Nframe=y, Nmarks=n, ExtNL=y, NLdist=1, Nh=6, Nw=6, Nmr=3](n2)(17,14){}
\node[Nframe=y, Nmarks=n, ExtNL=y, NLdist=1, Nh=6, Nw=6, Nmr=3](n3)(25,15){}
\node[Nframe=y, Nmarks=n, ExtNL=y, NLdist=1, Nh=6, Nw=6, Nmr=3](n4)(29,23){}

\node[Nframe=y, Nmarks=n, Nh=20, Nw=30, Nmr=0](M)(20,20){}
\node[Nframe=n, Nmarks=n](label)(8,33){$\M$}

\node[Nframe=n, Nmarks=n, Nh=0, Nw=0, Nmr=0](e1)(35,27){}
\node[Nframe=n, Nmarks=n, Nh=0, Nw=0, Nmr=0](e2)(35,20){}
\node[Nframe=n, Nmarks=n, Nh=0, Nw=0, Nmr=0](e3)(35,13){}

\node[Nframe=y, Nmarks=n, ExtNL=y, NLdist=1, Nh=12, Nw=12, Nmr=6](v1)(50,20){$v_{<0}$}
\nodelabel[ExtNL=n](v1){$-\frac{n^2 \cdot W}{\alpha^n}$}




\drawedge[ELpos=50, ELside=l, eyo=1, curvedepth=0](e1,v1){}
\drawedge[ELpos=25, ELside=l, curvedepth=0](e2,v1){{\scriptsize {\sf reset}}}
\drawedge[ELpos=50, ELside=l, eyo=-1, curvedepth=0](e3,v1){}

\drawline[AHnb=1,arcradius=1](50,14)(50,4)(3,4)(3,20)(8,20)
\node[Nframe=n, Nmarks=n](label)(30,6){{\scriptsize $A \cup \{{\sf reset}\}$}}








\end{picture}

%% file: figures/mdp-pos-neg.tex

\begin{picture}(60,50)(0,2)

\gasset{Nw=8,Nh=8,Nmr=4, rdist=1, loopdiam=5}

\node[Nframe=y, Nmarks=i](n0)(40,37){$-1$}
\nodelabel[ExtNL=y, NLangle=155, NLdist=2](n0){$\frac{1}{2}$}
\nodelabel[ExtNL=y, NLangle=0, NLdist=1](n0){$v_0$}
\node[Nframe=y, Nmarks=i](n1a)(10,17){$2$}
\nodelabel[ExtNL=y, NLangle=155, NLdist=2](n1a){$\frac{1}{2}$}
\nodelabel[ExtNL=y, NLangle=270, NLdist=1](n1a){$v_1$}
\node[Nframe=y](n1b)(30,17){$0$}
\nodelabel[ExtNL=y, NLangle=270, NLdist=1](n1b){$v_2$}
\node[Nframe=y](n2)(50,17){$-2$}
\nodelabel[ExtNL=y, NLangle=0, NLdist=1](n2){$v_3$}

\drawloop[ELside=l,loopCW=y, loopdiam=6](n0){$a$}
\drawloop[ELside=r,loopCW=n, loopdiam=6, loopangle=270](n2){$a,b$}

\drawloop[ELpos=70, ELside=r,loopCW=n, loopdiam=6, loopangle=304](n0){$\frac{1}{8}$}
\drawloop[ELpos=52, ELside=l,loopCW=n, loopdiam=6, loopangle=304](n0){$b$}



\drawedge[ELpos=50, ELside=l, curvedepth=6](n1a,n1b){$a,b$}
\drawedge[ELpos=50, ELside=l, curvedepth=6](n1b,n1a){$a$}
\drawedge[ELpos=50, ELside=r, curvedepth=0](n1b,n2){$b$}

\drawbpedge[ELpos=45, ELside=r, syo=-3](n0,270,10,n1b,45,3){$\frac{1}{8}$}
\drawbpedge[ELpos=45, ELside=l, syo=-3](n0,270,13,n2,135,3){$\frac{3}{4}$}
\drawarc(40,33,4,260,294)







\end{picture}

%% file: figures/mdp-ec-bound.tex

\begin{picture}(75,50)(0,2)

\gasset{Nw=8,Nh=8,Nmr=4, rdist=1, loopdiam=5}

\node[Nframe=y, Nmarks=n](n0)(15,45){$0$}
\nodelabel[ExtNL=y, NLangle=0, NLdist=1](n0){$v_E$}

\node[Nframe=n](n1)(5,25){}
\node[Nframe=n](n2)(15,25){}
\node[Nframe=n](n3)(25,25){}

\node[Nframe=y, Nw=24,Nh=24,Nmr=5](ec)(15,20){\begin{tabular}{c}end-comp.\\with value~$\eta$\end{tabular}}

\node[Nframe=n](label)(15,5){$w_1$}

\drawedge[ELpos=30, ELside=l, curvedepth=0](n0,n1){}
\drawedge[ELpos=30, ELside=l, curvedepth=0](n0,n2){}
\drawedge[ELpos=30, ELside=l, curvedepth=0](n0,n3){}

\node[Nframe=n, Nmarks=n](n0)(37,20){$\Rightarrow$}

\node[Nframe=y, Nmarks=n](n0)(60,45){$B$}
\nodelabel[ExtNL=y, NLangle=0, NLdist=1](n0){$v_E$}

\node[Nframe=n](n1)(50,25){}
\node[Nframe=n](n2)(60,25){}
\node[Nframe=n](n3)(70,25){}

\node[Nframe=y, Nw=24,Nh=24,Nmr=5](ec)(60,20){\begin{tabular}{c}every\\vertex\\has weight~$\eta$\end{tabular}}

\node[Nframe=n](label)(60,5){$w'$}

\drawedge[ELpos=30, ELside=l, curvedepth=0](n0,n1){}
\drawedge[ELpos=30, ELside=l, curvedepth=0](n0,n2){}
\drawedge[ELpos=30, ELside=l, curvedepth=0](n0,n3){}










\end{picture}

%% file: figures/mdp-ec-bound-difference.tex

\begin{picture}(40,60)(0,2)

\gasset{Nw=8,Nh=8,Nmr=4, rdist=1, loopdiam=5}

\node[Nframe=y, Nmarks=n](n0)(20,55){$-B$}
\nodelabel[ExtNL=y, NLangle=0, NLdist=1](n0){}  

\node[Nframe=n](n1)(10,35){}
\node[Nframe=n](n2)(20,35){}
\node[Nframe=n](n3)(30,35){}

\node[Nframe=y, Nw=24,Nh=24,Nmr=5](ec)(20,30){\begin{tabular}{c}weight of $v$ \\is $w(v) - \eta$,\\for all $v$\end{tabular}}

\drawedge[ELpos=30, ELside=l, curvedepth=0](n0,n1){}
\drawedge[ELpos=30, ELside=l, curvedepth=0](n0,n2){}
\drawedge[ELpos=30, ELside=l, curvedepth=0](n0,n3){}

\node[Nframe=n](n1)(12,25){}
\node[Nframe=n](n2)(20,25){}
\node[Nframe=n](n3)(28,25){}




\node[Nframe=y](m2)(20,10){$0$}
\drawedge[ELpos=30, ELside=l, eyo=2, curvedepth=-2](n1,m2){}
\drawedge[ELpos=30, ELside=l, eyo=2, curvedepth=0](n2,m2){}
\drawedge[ELpos=30, ELside=l, eyo=2, curvedepth=2](n3,m2){}
\drawloop[ELside=l,loopCW=y, loopdiam=6, loopangle=0](m2){}










\end{picture}

%% file: figures/mdp-pure-memory-needed.tex
\begin{picture}(150,45)(0,0)

\gasset{Nw=8,Nh=8,Nmr=4, rdist=1, loopdiam=5}

\node[Nframe=y, Nmarks=i](n0)(10,27.5){$v_0$}
\node[Nframe=y, Nh=0, Nw=0, Nmr=0, Nmarks=n, ExtNL=y, NLdist=5, NLangle=0](ndummy)(20,27.5){$\frac{1}{2}$}


\node[Nframe=y, Nmarks=n](n1)(35,35){$1$}   
\node[Nframe=y, Nmarks=n](n2)(50,35){$-2$}   
\node[Nframe=y, Nmarks=n](n3)(65,35){$1$}   

\node[Nframe=y, Nmarks=n](n4)(35,20){$-1$}   
\node[Nframe=y, Nmarks=n](n5)(50,20){$2$}   
\node[Nframe=y, Nmarks=n](n6)(65,20){$-1$}   
\node[Nframe=y, Nmarks=n](n7)(80,20){$v_7$}
\node[Nframe=y, Nmarks=n](n8)(95,20){$-1$}   
\node[Nframe=y, Nmarks=n](n9)(110,20){$2$}   
\node[Nframe=y, Nmarks=n](n10)(125,20){$-1$}   
\node[Nframe=y, Nmarks=n](n11)(140,12.5){$v_{14}$}

\node[Nframe=y, Nmarks=n](n12)(95,5){$1$}   
\node[Nframe=y, Nmarks=n](n13)(110,5){$-2$}   
\node[Nframe=y, Nmarks=n](n14)(125,5){$1$}   






\drawedge[AHnb=0, ELpos=65, ELside=l, ELdist=1, curvedepth=0](n0,ndummy){$a$}
\drawedge[ELpos=31, ELside=r, ELdist=0.2, curvedepth=0](ndummy,n1){}
\drawedge[ELpos=31, ELside=r, ELdist=0.2, curvedepth=0](ndummy,n4){}
\drawarc(20,27.5,3,333.44,26.56)

\drawedge[ELpos=50, ELside=l, curvedepth=0](n1,n2){}
\drawedge[ELpos=50, ELside=l, curvedepth=0](n2,n3){}
\drawedge[ELpos=50, ELside=r, curvedepth=-12](n3,n0){}

\drawedge[ELpos=50, ELside=l, curvedepth=0](n4,n5){}
\drawedge[ELpos=50, ELside=l, curvedepth=0](n5,n6){}
\drawedge[ELpos=50, ELside=l, curvedepth=0](n6,n7){}
\drawedge[ELpos=50, ELside=l, curvedepth=0](n7,n8){}
\drawedge[ELpos=50, ELside=l, curvedepth=0](n8,n9){}
\drawedge[ELpos=50, ELside=l, curvedepth=0](n9,n10){}
\drawedge[ELpos=50, ELside=l, curvedepth=0](n10,n11){}

\drawbpedge[ELpos=30, ELside=r, curvedepth=8](n0,270,30,n12,180,50){$b$}

\drawedge[ELpos=50, ELside=l, curvedepth=0](n12,n13){}
\drawedge[ELpos=50, ELside=l, curvedepth=0](n13,n14){}
\drawedge[ELpos=50, ELside=l, curvedepth=0](n14,n11){}

\drawloop[ELside=l,loopCW=y](n11){}








\end{picture}